\documentclass[12pt,]{article}
\usepackage{amsmath}
\usepackage[left=1in,top=1in,right=1in,bottom=1in]{geometry}
\newcommand*{\authorfont}{\fontfamily{phv}\selectfont}
\usepackage[]{mathpazo}

  \usepackage[T1]{fontenc}
  \usepackage[utf8]{inputenc}

\usepackage{abstract}

\renewenvironment{abstract}
 {{%
    \setlength{\leftmargin}{0mm}
    \setlength{\rightmargin}{\leftmargin}%
  }%
  \relax}
 {\endlist}

\makeatletter
\def\@maketitle{%
  \newpage
  \let \footnote \thanks
    {\fontsize{18}{20}\selectfont\raggedright  \setlength{\parindent}{0pt} \@title \par}%
}
\makeatother

\usepackage{longtable,booktabs}

\usepackage{graphicx,grffile}
\makeatletter
\def\maxwidth{\ifdim\Gin@nat@width>\linewidth\linewidth\else\Gin@nat@width\fi}
\def\maxheight{\ifdim\Gin@nat@height>\textheight\textheight\else\Gin@nat@height\fi}
\makeatother
\setkeys{Gin}{width=\maxwidth,height=\maxheight,keepaspectratio}

\title{The Increased Effect of Elections and Changing Prime Ministers on Topics Discussed in the Australian Federal Parliament between 1901 and 2018 \thanks{\textbf{Contact:} Comments, suggestions, and criticism of the 17 November 2021 version of this paper would be very welcome at: rohan.alexander@ utoronto.ca. \textbf{Acknowledgements:} Thank you to Chris Cochrane, Dan Simpson, Edward Morgan, Evan Roberts, Hugo Geils, Jill Sheppard, John McAndrews, John Tang, Leslie Root, Martine Mariotti, Matt Jacob, Matthew Kerby, Myles Clark, Ruth Howlett, Ryan Briggs, Tianyi Wang, Tim Hatton, and Zach Ward for their helpful suggestions; and to the UC Berkeley Demography Department for the use of their computing resources. We are grateful for the many excellent comments that we received from seminar participants at the Australian National University (ANU) SPIR, the ANU RSE, the Australian Parliamentary Library, the Max Planck Institute for Demographic Research, the University of Toronto Political Behavior Group, and the 2019 PSA Political Methodology Conference. \textbf{Data availability statement:} The replication materials for this paper can be found at: \url{https://github.com/RohanAlexander/hansard}. \textbf{Conflicts of interest:} There are no conflicts of interest to disclose. \textbf{Author contributions:} RA had the original idea, created the datasets, and did the topic modelling. MA built the analysis model. Both RA and MA analysed and interpreted the data, contributed to writing the paper, and approved the final version.}  }

\author{\Large Rohan Alexander\vspace{0.05in} \newline\normalsize\emph{University of Toronto}   \and \Large Monica Alexander\vspace{0.05in} \newline\normalsize\emph{University of Toronto}  }

\date{}

\usepackage{titlesec}

\titleformat*{\section}{\normalsize\bfseries}
\titleformat*{\subsection}{\normalsize\itshape}
\titleformat*{\subsubsection}{\normalsize\itshape}
\titleformat*{\paragraph}{\normalsize\itshape}
\titleformat*{\subparagraph}{\normalsize\itshape}

\usepackage{natbib}
\usepackage[strings]{underscore} 

\usepackage{setspace}

\makeatletter
\@ifpackageloaded{hyperref}{}{%
\ifxetex
  \PassOptionsToPackage{hyphens}{url}\usepackage[setpagesize=false, 
              unicode=false, 
              xetex]{hyperref}
\else
  \PassOptionsToPackage{hyphens}{url}\usepackage[unicode=true]{hyperref}
\fi
}

\@ifpackageloaded{color}{
    \PassOptionsToPackage{usenames,dvipsnames}{color}
}{%
    \usepackage[usenames,dvipsnames]{color}
}
\makeatother
\hypersetup{breaklinks=true,
            bookmarks=true,
            pdfauthor={Rohan Alexander (University of Toronto) and Monica Alexander (University of Toronto)},
             pdfkeywords = {text-as-data, Australian politics, unsupervised machine learning, Bayesian hierarchical Dirichlet model, Comparative Agendas Project},  
            pdftitle={The Increased Effect of Elections and Changing Prime Ministers on Topics Discussed in the Australian Federal Parliament between 1901 and 2018},
            colorlinks=true,
            citecolor=blue,
            urlcolor=blue,
            linkcolor=magenta,
            pdfborder={0 0 0}}
\makeatletter
\def\fps@figure{htbp}
\makeatother

\usepackage{booktabs}
\usepackage{longtable}
\usepackage{array}
\usepackage{multirow}
\usepackage{wrapfig}
\usepackage{float}
\usepackage{colortbl}
\usepackage{pdflscape}
\usepackage{tabu}
\usepackage{threeparttable}
\usepackage{threeparttablex}
\usepackage[normalem]{ulem}
\usepackage{makecell}
\usepackage{subfig}
\usepackage{booktabs}
\usepackage{longtable}
\usepackage{array}
\usepackage{multirow}
\usepackage{wrapfig}
\usepackage{float}
\usepackage{colortbl}
\usepackage{pdflscape}
\usepackage{tabu}
\usepackage{threeparttable}
\usepackage{threeparttablex}
\usepackage[normalem]{ulem}
\usepackage{makecell}
\usepackage{xcolor}

\providecommand{\tightlist}{%
\setlength{\itemsep}{0pt}\setlength{\parskip}{0pt}}

\begin{document}
	
%


{
\setlength{\parindent}{0pt}
\thispagestyle{plain}
{\fontsize{18}{20}\selectfont\raggedright 
\maketitle  

}

{
   \vskip 13.5pt\relax \normalsize\fontsize{11}{12} 
\textbf{\authorfont Rohan Alexander} \hskip 15pt \emph{\small University of Toronto}   \par \textbf{\authorfont Monica Alexander} \hskip 15pt \emph{\small University of Toronto}   

}

}

\begin{abstract}

    \hbox{\vrule height .2pt width 39.14pc}

    \vskip 8.5pt 

\noindent Politics and discussion in parliament is likely to be influenced by the party in power and associated election cycles. However, little is known about the extent to which these events affect discussion and how this has changed over time. We systematically analyse how discussion in the Australian Federal Parliament changes in response to two types of political events: elections and changed prime ministers. We use a newly constructed dataset of what was said in the Australian Federal Parliament from 1901 through to 2018 based on extracting and cleaning available public records. We reduce the dimensionality of discussion in this dataset by using a correlated topic model to obtain a set of comparable topics over time. We then relate those topics to the Comparative Agendas Project, and then analyse the effect of these two types of events using a Bayesian hierarchical Dirichlet model. We find that: changes in prime minister tend to be associated with topic changes even when the party in power does not change; and the effect of elections has been increasing since the 1980s, regardless of whether the election results in a change of prime minister.

 \par

    \hbox{\vrule height .2pt width 39.14pc}

\end{abstract}

\vskip 6.5pt

\noindent \singlespacing \hypertarget{introduction}{%
\section{Introduction}\label{introduction}}

What is discussed in parliament is a key indicator of a government's priorities and likely policy outcomes. While in an ideal world, we may hope that topics of discussion represent the most important social, economic, and environmental issues for the country at that particular time, in reality, discussion is heavily influenced by party politics, the priorities of the government of the day, and the current election cycle.

While there is evidence to suggest a relationship between parliamentary discussion and political events, little research exists that systematically analyses the sensitivity of topics discussed in parliament to political changes in a statistical way. This is mostly likely because historical data are difficult to obtain, and parliamentary discourse and events are inherently complex and multi-dimensional, and difficult to operationalise into an analysable form. However, the increased accessibility of optical character recognition to parse historical documents, and recent development in statistical text analysis, allow for a broader analysis of parliamentary events and discussion over time.

In this paper, we examine the effect of elections and changes in prime ministers on the topics that are discussed in the Australian Federal Parliament. We construct a new dataset from the text record, known as `Hansard', of what was said in the Australian Federal Parliament since its inception in 1908, to 2018. The dataset covers records for 7,934 days in the House of Representatives (lower house) and 6,746 days in the Senate (upper house).

We then systematically analyse the text contained in the Hansard dataset in two stages, each of which takes advantage of a different statistical technique. First, we used a correlated topic model to obtain a set of comparable topics discussed in parliament over time, and then further reduce those topics to match the Comparative Agendas Project. Second, we introduce a new modeling approach to examine the association between changes in topics discussed changes in prime minister and elections. The second modeling approach centers on a Bayesian hierarchical Dirichlet model, which allows for the effect of these two event types to be modeled concurrently, while taking time since election into account. The model also allows for information about event effects to be pooled across documents.

We find that: firstly, changes in prime minister tend be associated with topic changes even when the party in power does not change. For instance, the change from Hughes to Bruce in 1923, Forde to Chifley in 1945; Menzies to Holt in 1966; Hawke to Keating in 1991; and Rudd to Gillard in 2010 are all associated with significant changes in topics despite no change in the party in power. Secondly, as expected, elections where the party in power also changes, such as Fisher in 1910 and 1914, Menzies in 1949, Hawke in 1983, and Howard in 1996 are associated with topic changes, but the 1974 Whitlam, and 1984 Hawke re-elections stand out as elections where the prime minister did not change but there was a significant change in topics.

As our dataset covers 118 years we are able to see how the effect of these two types of events changes over time. We find that in Australia the effect of elections and a change in prime minister appear to have become more pronounced since the 1980s. With a small number of exceptions, in the first half of our dataset, even changes in prime minister where the party in power also changed were not associated with overly large changes in the topics of parliamentary discussion. It may be that more recent prime ministers are trying to more thoroughly distinguish themselves from their predecessor, or that the role of the government in agenda setting in the Australian Federal Parliament has changed.

Our work contributes to the growing literature that analyses text using quantitative methods. It sits across, and draws from, various historically-separate disciplines including applied statistics, economics, and political science. In addition to our findings about the effect of events in Australian politics, we contribute to this literature in terms of both data and methods. From a data perspective we bring to bear an essentially-complete record of what was said in the Australian Federal Parliament on a daily basis, and our dataset is available to other researchers. From a methods perspective, we present a modeling framework that can be generalizable to other problems looking at the association between text data and events. The model allows for time since event to be taken into account; implements pooling across groups of similar documents; and additionally identifies potential outlying topic distributions without the need to pre-specify the event of interest. There are many avenues for closely related future work such as: investigating the effect of other types of events; including a richer set of covariates to disentangle the reason for different effects; and reversing the causality to examine the effect of what is said in parliament on various outcomes.

The remainder of the paper is structured as follows. Section \ref{hansarddata} examines the data that we use; Section \ref{model} introduces our models; Section \ref{results} goes into our results; and finally, Section \ref{discussion} discusses our results and some of the shortcomings.

\hypertarget{hansarddata}{%
\section{Data}\label{hansarddata}}

The first step is to construct a dataset derived from the Australian Hansard, which contains the text of what was said throughout history in the Australian Federal Parliament. This section gives on the availability and characteristics of such data and outlines the method and steps undertaken to create a ready-to-analyse dataset.

\hypertarget{background}{%
\subsection{Background}\label{background}}

The term `Hansard' refers to a daily text record of what was said in parliament. The UK, Australia and Canada all have a Hansard, which span historical periods right up to present day. Following the example of other countries, the Hansard for the Australian Federal Parliament has been made available since Federation in 1901. Analysing Hansard records and their equivalents is increasingly viable as new methods and reduced computational costs make it easier. While Hansard is not necessarily verbatim, it is considered close enough for text-as-data purposes. For instance, \citet{Mollin2008} found that in the case of the UK Hansard the differences would only affect specialised linguistic analysis. \citet{Edwards2016} examined Australia, New Zealand and the UK, and found that changes were usually made by those responsible for creating the Hansard record, instead of the parliamentarians. As those who create Hansard are tasked with creating an accurate record of proceedings, this suggests the records should be fit for the purpose of our analysis.

The recent digitisation of Hansard records has allowed increased analysis of parliament text records. For instance, \citet{RheaultCochran2018} examined ideology and party polarisation in Britain and Canada using word embeddings. In the UK, \citet{Duthie2016} examined which politicians made supportive or aggressive statements toward other politicians, and \citet{PetersonSpirling2018} examined polarisation. One exciting aspect of research using the UK Hansard has been linking text records with other datasets. For instance, \citet{SlapinKirklandLazzaroLeslie2018} linked votes and speeches to examine grandstanding within parties. As digitisation methods improve, increasingly older UK records can be analysed, for instance, \citet{Dimitruk2018} considered the effect of estate bills on prorogations in seventeenth century England. In New Zealand, \citet{Curran2018} modelled the topics discussed between 2003 and 2016, and \citet{Graham2016} examined unparliamentary language between 1890 and 1950.

Parts of Australian Hansard records have been analysed for various purposes and our paper contributes to a small but growing literature. For instance, \citet{Rasiah2010} examined Hansard records for the Australian House of Representatives to examine whether politicians attempted to evade questions about Iraq during February and March 2003. \citet{GansLeigh2012} examined Australian Hansard records to associate mentions by politicians of certain public intellectuals with neutral or positive sentiment. \citet{Salisbury2011} examined unparliamentary behaviour. And \citet{FraussenGrahamHalpin2019} examined Australian Hansard records to assess the prominence of interest groups. The closest research to ours that we have found is \citet{Boulous2013} who examined parliamentary debate in Australia between 1946 and 2012. To our knowledge, this is the first paper that analyses the complete Hansard record since 1901.

\hypertarget{creation-of-hansard-dataset}{%
\subsection{Creation of Hansard dataset}\label{creation-of-hansard-dataset}}

The Australian Federal Parliament makes daily Hansard records available online as PDFs and these are considered the official release.\footnote{Although we do not use them here, XML records are also available in most cases. Tim Sherratt makes Commonwealth XML records for 1901 to 1980 available as a single download at: \url{http://historichansard.net/}. Commonwealth XML records from 1998 to 2014 are available from Andrew Turpin's website, and from 2006 through to today from Open Australia's website. The records can also be downloaded from the Australian Hansard website or the website can be scraped. We do not use the XML records or scrape the website for this paper because those records are known to be incomplete but the extent of how incomplete they are is unknown. The trade-off for a more-complete record is the errors introduced by having to parse the PDFs.} We provide an example of a Hansard PDF page in Appendix \ref{examplehansardpage}. There are 14,680 PDF files that cover the Hansard over the entire period 1901-2018. Our goal is to take these PDFs and convert them into a digitized dataset of text that is able to be analysed.

The creation of the Hansard dataset involves the following steps. Firstly, optical character recognition (OCR) is used to convert static PDFs into digitised text. This creates a text file for each PDF with a single character string. Secondly, as many of the original PDFs contained two columns per page, the text file is reshuffled to ensure the text is in the right order. Thirdly, the text is split out such that it is separated by speaker. This involves identifying names and the structure of how speakers are introduced and then splitting based on this. Lastly, the dataset is created in tidy format \citep{wickham2014tidy}, such that there are columns for speaker, the date, and the text of what was said, and every row refers to a different speaker.

These steps were performed using R \citep{R2018}.\footnote{Our code and data are available on request or via the GitHub repository for this paper: \url{https://github.com/RohanAlexander/hansard}. The scripts are primarily based on: the \texttt{PDFtools} R package of \citet{Ooms2018pdftools}; the \texttt{tidyverse} R package of \citet{Wickham2017}; the \texttt{tm} R package of \citet{FeinererHornik2018}; the \texttt{lubridate} R package of \citet{GrolemundWickham2011}; the \texttt{tidytext} R package of \citet{SilgeRobinson2016}; and the \texttt{stringi} R package of \citet{Gagolewski2018}. The functions of those packages are augmented by: the \texttt{furrr} R package of \citet{VaughanDancho2018}; and the \texttt{tictoc} R package of \citet{Izrailev2014}. The \texttt{hunspell} R package of \citet{Ooms2017} is used to help find spelling issues; and the \texttt{quanteda} R package of \citet{Benoit2018} is used to compound multiword expressions.} Some error is introduced at this stage because many of the records are in a two-column format that need to be separated, and the PDF parsing is not always accurate, especially for older records. An example of the latter issue is that `the' is often parsed as `thc'. These errors are corrected when they can be identified, but as there are almost a billion words in the dataset, we are restricted to changes that can be made at scale.

There are 14,680 days of publicly available Hansard records across the two chambers of the Australian Federal Parliament for which we have PDFs. Further summary statistics for this are provided in Appendix \ref{hansardsummarystatistics}. In general, the frequency of sitting days based on information from the Hansard dataset compared to the actual number of sitting days is comparable, almost complete coverage of our dataset.

Our data cleaning process indicates concerns with a small number of PDFs and these are detailed in Appendix \ref{knownhansardissues}. The percentage of stop-words each day is reasonably consistent over time (see Appendix \ref{stopwordsgraph}). This suggests that the data are fit-for-purpose, although manual inspection does suggest there is some improvement in quality over time.

\hypertarget{creation-of-analysis-dataset}{%
\subsection{Creation of analysis dataset}\label{creation-of-analysis-dataset}}

Using the Hansard dataset, we pre-process the text to create an analysis dataset to model topics and to subsequently investigate the relationship between topics and events. The specific steps that we take are to: remove numbers and punctuation; change the words to lower case; and concatenate multi-word names titles and phrases, such as new south wales to new\_south\_wales. Then the sentences are de-constructed and each word considered individually. We do not stem the words because, following \citet{SchofieldMimno2016}, we were not able to see much appreciable benefit. The resulting dataset used for analysis contains counts of words by day for 14,680 sitting days between 1901 and 2018.\footnote{Our dataset is available at: \url{https://github.com/RohanAlexander/hansard}. While we are making our data public in an attempt to help other researchers, we cleaned the dataset toward the requirements of this paper.}

\hypertarget{model}{%
\section{Model}\label{model}}

The goal of our modelling strategy is twofold. Firstly, we want to use topic modelling \citep{Blei2003latent} to summarise the Hansard text into meaningful topics that reduce the dimensionality of the text data and capture the main themes discussed in parliament over time. Secondly, we want to relate the resulting topic distributions to elections and changing prime ministers over time, accounting for temporal trends.

We first use a Correlated Topic Model \citep{BleiLafferty2007} to obtain estimated topic distributions over time, and then further group these topics to match those of the Comparative Agendas Project. We consider these topic distributions as inputs that can be analysed by another model. Thus, the second modelling step involves using a Bayesian hierarchical Dirichlet model to analyse changes in the topic distributions (obtained from the first step) in relation to events of interest.

In the following section, we briefly describe the topic modelling approach, before discussing the Bayesian hierarchical Dirichlet analysis model used to investigate changes in topics. Background detail on topic modelling is available in Appendix \ref{LDAexample}.

\hypertarget{overview-of-topic-modelling-and-topic-selection}{%
\subsection{Overview of topic modelling and topic selection}\label{overview-of-topic-modelling-and-topic-selection}}

Although more- or less-fine levels of analysis are possible, here we are primarily interested in considering a day's topics. This means that each day's Hansard record needs to be classified by its topics. Sometimes Hansard records includes titles that make the topic clear. But not every statement has a title and the titles do not always define topics in a well-defined and consistent way, especially over longer time periods.

Other work such as \citet{BaumgartnerJones1993} and \citet{DowdingHindmoor2010} addressed this problem by creating a standardised codebook of policy categories and sub-categories and then manually assigning text to topics as appropriate. This approach ensures the categorisation is reasonable but as it is a manual process the size of the text that can be categorised is limited. \citet{AshMorelliOsnabru2018} combined the best of both in the context of the New Zealand Hansard, using a supervised machine learning approach that would likely be beneficial in the Australian Hansard case as well.

In order to effectively categorise the topics of the entire Hansard, we use topic modelling, a statistical technique which aims to extract the underlying or latent `topics' from a collection of texts. In particular, we use a method which is similar to Latent Dirichlet Allocation (LDA), first developed by \citet{Blei2003latent}.

The key assumption behind LDA is that each day's text, `a document', in Hansard is made by speakers who decide the topics they would like to talk about in that document, and then choose words, `terms', that are appropriate to those topics. A topic could be thought of as a collection of terms, and a document as a collection of topics, where these collections are defined by probability distributions. The topics are not specified \emph{ex ante}; they are an outcome of the method, and it is in this sense that this approach can be considered unsupervised machine learning. Terms are not necessarily unique to a particular topic, and a document could be about more than one topic. The goal is to have the words found in each day's Hansard group themselves to define topics. This can provide more flexibility than other approaches such as a strict word count method, but can require a larger dataset and make interpretation more difficult.

An overview, and an example, of how topic modelling works is available in Appendix \ref{LDAoverviewandexample}. The underlying document generation process is discussed in Appendix \ref{LDAdocgenprocess}, and then Appendix \ref{LDAposteriorestimation} explains how this is reversed to generate topics given documents.

One notable limitation of LDA is that the model assumes that the presence of one topic is not correlated with the presence of another topic. However, in reality topics are often related. For instance, in the Hansard context, we may expect topics related to the army to be more commonly found with topics related to the navy, but less commonly with topics related to banking. As such, we use the Correlated Topic Model (CTM) of \citet{BleiLafferty2007} to obtain topic distributions. The CTM is a modification of LDA that allows for correlations between topics. More detail about the CTM is in Appendix \ref{correlatedtopicmodelsection}.

\hypertarget{categorisation-of-topics-using-the-comparative-agendas-project}{%
\subsubsection{Categorisation of topics using the Comparative Agendas Project}\label{categorisation-of-topics-using-the-comparative-agendas-project}}

There are various methods to help determine the appropriate number of topics to specify. Appendix \ref{selecttopicnumber} contains a discussion of how the number of topics was chosen based on the output of the CTM. However, we found that the optimal number of topics selected was too large to be tractable by our analysis model, and too large to be easily visualised, or examined. In addition, many of the topics in the optimal set referred to similar themes, such as budgeting or administrative topics.

For this reason we manually group the topics identified by the CTM into those of the Comparative Agendas Project. The Comparative Agendas Project provides a classification of a large number of specific issues into a small number of major topics.

\hypertarget{analysismodelexplanation}{%
\subsection{Analysis model}\label{analysismodelexplanation}}

The output of interest from the topic modelling stage is the proportion of each topic appearing in each document. The aim of the analysis stage of the modelling process is to analyse how the distribution of topics changes in relation to different types of events. But with many topics for each of the roughly 14,680 chamber-sitting-days spanning 118 years, the data are still too noisy to easily visualise changes around events.

One option for relating the topic distributions to events would be to use the Structural Topic Model (STM) of \citet{RobertsStewartAiroldi2016}. The distinguishing aspect of the STM is that it considers more than just a document's content when constructing topics. For example, we may believe a document's author, or, in the case of our paper, the prime minister or election period, may affect the topics within that document. The STM allows this additional information, or metadata, to affect the construction of topics, though influencing either topical prevalence or topical content. The assumption that there is some document generation process is the same as in LDA, it is just that this process now includes metadata.

However, the STM covariate framework has several limitations in terms of our goal to assess the relationship between topics and events:

\begin{enumerate}
\def\labelenumi{\arabic{enumi}.}
\tightlist
\item
  There is no way of specifying more complicated auto-correlated functional forms of the effects of events over time. For example, we believe that the effect of an election would peak at the time of the election, then gradually decay as a function of days since election. In the STM framework, it is possible to specify a constant or linear effect of elections over time, or a spline relationship over elections, but it is not possible to restrict the effect of a specific election over time to be monotonically decreasing. We are interested in time effects within election periods, which requires a more flexible framework.
\item
  There is no way to implement partial pooling across groups of similar documents. The STM framework assumes that documents are independently and identically distributed, conditional on the model covariates. However, it could be expected that topic distributions within a particular prime minister's time, for example, may be more- or less-likely to contain certain topics for reasons that are not reflected in the topic prevalence covariates. To account for this, we would like a covariate model that allows for the partial pooling of variance in topic distributions by group, such as sitting period, or election period.
\item
  There is no way of identifying `outlying' topic distributions -- and therefore events that had an important effect -- without pre-specifying the event of interest in the model. For example, if we think that the 9/11 attacks had an effect on parliamentary discourse, then a dummy for 9/11 would have to be included in the STM framework, but the specifics of the dummy construction affect the results. Instead we would like to identify important events based on different-to-expected topic distributions, after accounting for prime minister and election effects.
\end{enumerate}

To overcome these challenges, we formalise a statistical framework that allows us to systematically identify significant changes in topic distributions over time. Specifically, we use the estimated topic distributions from the previous step as an input into a Bayesian hierarchical Dirichlet regression framework, which relates the proportions of each topic to underlying time trends, changes in prime minister and elections.

\hypertarget{model-set-up}{%
\subsubsection{Model set-up}\label{model-set-up}}

Define \(\theta_{c,d,p}\) to be the proportion of speech that refers to topic \(p\) on day \(d\) in chamber \(c\) (where chamber refers to either the House of Representatives or the Senate). Note that the \(\theta_{c,d,1:P}\) for \(p = 1,2,\dots, P = 19\) are equal to the estimated values of \(\theta_{cd}\) from the CTM after we group them into the Comparative Agendas Project topics. We assume that the majority of variation in topics by day \(d\) is across sitting periods, \(s\), where a sitting period is defined as any group of days that are less than one week apart. Using this definition, there are 822 sitting periods over the period 1901 to 2018 inclusive. Appendix \ref{hansardsummarystatistics} contains more information about the sitting patterns over the course of the year, which have changed considerably since Federation.

The topic proportions on day \(d\) in chamber \(c\) are modelled in reference to their membership of a particular sitting period \(s\). Firstly, we assume that each distribution of topics, \(\theta_{c,d,1:P}\) for each day is a draw from a Dirichlet distribution with mean parameter \(\mu_{c,s[d],1:P}\):
\[
\theta_{c,d,1:P} \sim \mbox{Dirichlet}(\mu_{c,s[d],1:P})
\]
where the notation \(s[d]\) refers to the sitting period \(s\) to which day \(d\) of chamber \(c\) belongs. This distributional assumption accounts for the fact that on any given day in either chamber, the sum of all proportions in each topic must be one.

The goal of the model is to relate these proportions to the current prime minister \(g\) and election period \(e\), assuming that an `election effect' would be at its peak in the days close to the election, then decay over the period. The mean parameters \(\mu_{c,s,p}\) are modelled on the log scale as:

\[
\log \mu_{c,s,p} = \alpha_{g[c,s],p} + \beta_{e[c,s],p} \cdot (N_{s[e]} - s) + \delta_{c,s,p}
\]
where: \(\alpha_{g[c,s],p}\) is the mean effect for prime ministers \(g\) (which covers sitting period \(s\) in chamber \(c\)) and topic \(p\); \(\beta_{e[c,s],d,p}\) is the effect of election \(e\) (which occurs in sitting period \(s\), in chamber \(c\)) for topic \(p\); \(N_s\) is the total number of sitting periods in election period \(e\); and \(\delta_{c,s,p}\) is a random, or levels, effect for each chamber, sitting period and topic.

The term for the prime minister, \(\alpha_{g[c,s],p}\), assumes there is some underlying mean effect of each prime minister on the topic distribution that is common across chambers. Non-informative priors are placed on \(\alpha_{g[c,s],p}\):

\[
\alpha_{g[c,s],p} \sim \mbox{Normal}(0, 100).
\]

The election term, \(\beta_{e[c,s],p}\), assumes there is an initial effect of an election on the topic distribution, which then decays as a function of sitting periods since election, \(s\). In the model above, \(\beta_{e[c,s],p}\) is multiplied by the numbers of sitting periods since the election, \((N_{s[e]} - s)\). When the sitting period counter is small, \((N_{s[e]} - s)\) is large, so the election effect is larger.

One advantage of our model over using the STM is that we can restrict the effect of an election to be monotonically decreasing. This allows us to identify differences between prime minister and election effects even when there is a one-term prime minister. The value of the initial effect, \(\beta_{e[c,s],p}\), has a non-informative prior:
\[
\beta_{e[c,s],p} \sim \mbox{Normal}(0, 100).
\]

Finally, the sitting-period-specific random effect \(\delta_{c,s,p}\) allows for chamber- and sitting-period specific deviations in the topic distributions. This allows us to identify periods where there may be differences in the topics discussed across the two chambers. It also allows us to identify large deviations away from the expected distribution, thus helping to identify the effect of other, non-prime-minister and non-election events. In addition, this set up also partially pools effects across sitting periods. The \(\delta_{c,s,p}\) values are modelled hierarchically as:

\[
\delta_{c,s,p} \sim \mbox{Normal}(\mu_{c,p}, \sigma_{g[c,s],p}^2).
\]
This structure assumes that the sitting period effect is a draw from a distribution with a mean that is common to that particular chamber and topic, with some associated variance. This allows for the sitting-period-specific random effect \(\delta_{c,s,p}\) to be partially informed by other sitting periods for that chamber and topic combination.

The variance parameters \(\sigma_{g[c,s],p}^2\) give an indication of the how the variation in topics is changing over government periods. If the estimates of the variance are larger, then there is more variation in the topics discussed within a government period. Non-informative priors are placed on the variance parameters:

\[
\sigma_{g[c,s],p} \sim \mbox{Uniform}(0,3).
\]

We run the model in JAGS using the \texttt{rjags} package of \citet{Plummer2018}.

\hypertarget{results}{%
\section{Results}\label{results}}

Firstly, we describe the results of the topic modeling and aggregation using the CTM approach, which defined 80 topics over the period 1901 to 2018,which we then further reduced to match the Comparative Agendas Project. We then describe the results of the Bayesian analysis model, which identified prime ministers, elections and other events that were associated with a change in the topics discussed.

\hypertarget{topic-modelling}{%
\subsection{Topic modelling}\label{topic-modelling}}

We applied the CTM approach on the processed Hansard text database outlined in Section \ref{hansarddata}. The main output of interest are the types of topics identified by the model, and the prevalence of each topic for each day of parliamentary discussion.

Our main results are based on a topic model with 80 distinct topics. With almost 15,000 days and 80 topics, the analysis model is being fit to more than a million observations. The choice of 80 topics was made as a trade-off between standard diagnostic tests that suggested a larger number of topics would be more appropriate, and the need for the analysis model to be tractable. Those diagnostic tests are detailed in Appendix \ref{selecttopicnumber}.

LDA output defines a topic as a distribution of probabilities over words. In Table \ref{tab:topwordscomp} we display the ten words with the highest association for each of the 80 topics, but the topics are defined by probability distributions over all words. LDA does not apply labels to each topic, or collection of words, instead this has to be done by inspection. The topics cover areas such as budgets, demography, transport and infrastructure, war and conflict, health, education, agriculture, and trade. Similar to when topic models are run using the parliamentary text records of other countries, there are also some topics that seem to be about procedural or day-to-day matters, such as Topics 7 or 9. As expected, some topics seem to somewhat overlap with their content: for instance, Topics 4, 26, 28, 30 and 66 all relate to war and conflict.

In the final column of Table \ref{tab:topwordscomp} we summarise the results of our 80-topic model using the categories of the Comparative Agendas Project (CAP), as summarised by \citet{Bevan2017}.\footnote{We use Version 1.0 of the Master Codebook, as at 31 July 2014 available at \url{http://sbevan.com/cap-master-codebook.html}.} We do this firstly, to assist other researchers in understanding our results, by trying to relate them to a more well-known approach. We also do this because it allows us to further reduce our 80-topic model to only 21 topics.\footnote{Although the CAP Codebook topic numbering goes to 23, there are only 21 topics because there is no topic 11 or topic 22. Our 80 topics did not relate to CAP topics 21 or 23, meaning that we have 19 CAP-based categories.} It is these 19 CAP groupings to which we apply our analysis model.

\begingroup\fontsize{7}{9}\selectfont

\begin{longtable}[t]{rll}
\caption{\label{tab:topwordscomp}The ten words most strongly associated with each topic}\\
\toprule
Topic & Terms & CAP name\\
\midrule
\endfirsthead
\caption[]{\label{tab:topwordscomp}The ten words most strongly associated with each topic \textit{(continued)}}\\
\toprule
Topic & Terms & CAP name\\
\midrule
\endhead

\endfoot
\bottomrule
\endlastfoot
1 & women, rights, marriage, human, discrimination, law, equal, community, society, support & Civil Rights\\
2 & death, compensation, injury, estate, abolition, accident, injured, deaths, loss, died & Labor\\
3 & constitution, parliament, power, powers, constitutional, referendum, convention, representatives, proposal & Government Operations\\
4 & defence, forces, personnel, army, military, defence\_force, equipment, base, aircraft, air & Defense\\
5 & party, communist, matter, time, mckenna, communists, organization, country, position, henty & Civil Rights\\
\addlinespace
6 & vietnam, countries, south, china, united\_states, world, aid, asia, country, foreign & International Affairs\\
7 & petition, petitioners, citizens, pray, parliament, assembled, representatives, duty, bound, undersigned & Government Operations\\
8 & sugar, industry, bounty, queensland, growers, production, fruit, cotton, ton, paid & Agriculture\\
9 & na, senate, president, question, greens, time, committee, support, australians, country & Government Operations\\
10 & service, public, board, officers, officer, department, salary, commissioner, appointment, salaries & Labor\\
\addlinespace
11 & senators, ill, time, gardiner, measure, western\_australia, collings, position, read, leader & Government Operations\\
12 & television, broadcasting, service, stations, radio, post, services, commercial, abc, telephone & Technology\\
13 & senate, senators, chamber, representatives, representing, business, party, week, position, public & Government Operations\\
14 & workers, employees, relations, industrial, employers, workplace, employment, employer, union, business & Labor\\
15 & president, sympathy, public, word, world, personal, regret, presiding, standing, bias & Government Operations\\
\addlinespace
16 & commission, report, royal, royal\_commission, inquiry, evidence, commissioner, commissions, body & Government Operations\\
17 & tax, income, taxation, sales, treasurer, per\_cent, pay, revenue, rate, taxes & Macroeconomics\\
18 & fl, senate, question, greens, time, carbon, president, support, change, move & Environment\\
19 & industrial, union, arbitration, workers, unions, trade, court, industry, conciliation, employers & Labor\\
20 & matter, labor\_party, question, debate, situation, time, organisation, lo, cavanagh, greenwood & Government Operations\\
\addlinespace
21 & department, matter, time, question, regard, money, expenditure, connection, business, information & Domestic Commerce\\
22 & per\_cent, ad, val, subitem, item, omitting, inserting, exceeding, duty, intermediate & Foreign Trade\\
23 & court, law, high\_court, justice, federal, courts, attorneygeneral, legal, judge, tribunal & Law and Crime\\
24 & debate, time, labor\_party, issue, deputy, political, question, matter, country, process & Government Operations\\
25 & life, superannuation, fund, insurance, scheme, funds, national, contributions, retirement, age & Labor\\
\addlinespace
26 & security, iraq, support, detention, international, intelligence, time, world, australias, terrorism & Defense\\
27 & education, schools, students, school, university, universities, training, student, funding, children & Education\\
28 & war, production, country, matter, control, governments, industry, time, prices, services & Defense\\
29 & health, medical, hospital, private, insurance, medicare, hospitals, scheme, services, public & Health\\
30 & defence, military, naval, training, navy, forces, officers, time, force, service & Defense\\
\addlinespace
31 & matter, information, letter, department, evidence, document, documents, report, statement, office & Government Operations\\
32 & british, great\_britain, germany, empire, trade, country, canada, new\_zealand, imperial, conference & International Affairs\\
33 & committee, report, parliament, committees, public, parliamentary, recommendations, time, joint, inquiry & Government Operations\\
34 & immigration, country, migration, migrants, citizenship, policy, immigrants, citizens, english, countries & Immigration\\
35 & question, time, debate, standing\_orders, matter, business, parliament, standing, chairman, chair & Government Operations\\
\addlinespace
36 & tasmania, queensland, western\_australia, new\_south\_wales, south\_australia, victoria, federal, south & Government Operations\\
37 & rules, december, service, hon, association, january, november, october, july, june & Government Operations\\
38 & housing, building, homes, home, capital, site, houses, new\_south\_wales, construction, canberra & Housing\\
39 & question, department, answer, notice, provided, services, total, staff, nil, ii & Government Operations\\
40 & shipping, ships, ship, vessels, line, trade, port, vessel, ports, sea & Foreign Trade\\
\addlinespace
41 & amendments, subsection, section, schedule, omit, item, line, substitute, person, title & Government Operations\\
42 & prime\_minister, party, country, leader, parliament, policy, opposite, political, time, election & Government Operations\\
43 & aircraft, aviation, air, airport, airlines, transport, civil, qantas, airline, services & Transportation\\
44 & duty, per\_cent, item, duties, imported, committee, revenue, industry, article, manufacturers & Foreign Trade\\
45 & main, electorate, million, committee, community, regional, services, per\_cent, time, program & Government Operations\\
\addlinespace
46 & late, parliament, loss, time, lost, public, memorial, friend, passing, regret & Government Operations\\
47 & report, senate, matter, democrats, governments, leave, per\_cent, aboriginal, program, button & Government Operations\\
48 & roads, road, water, railway, line, transport, construction, river, country, money & Transportation\\
49 & development, time, service, national, programme, overseas, field, matter, department, country & International Affairs\\
50 & world, nations, international, united\_nations, countries, treaty, peace, japan, united\_states, japanese & International Affairs\\
\addlinespace
51 & tariff, industry, trade, industries, board, customs, protection, country, duties, duty & Foreign Trade\\
52 & pension, pensions, pensioners, week, social, age, benefits, service, services, repatriation & Labor\\
53 & community, electorate, na, time, local, support, australians, day, ms, national & Government Operations\\
54 & northern\_territory, territory, regulations, regulation, parliament, council, governor\_general, ordinance & Government Operations\\
55 & ill, money, position, country, time, amount, financial, matter, treasurer, dr & Macroeconomics\\
\addlinespace
56 & environment, heritage, nsw, environmental, conservation, community, project, management, forest & Environment\\
57 & energy, gas, nuclear, fuel, change, industry, emissions, power, climate, carbon & Energy\\
58 & question, department, matter, answer, time, notice, information, report, questions, national & Government Operations\\
59 & per\_cent, budget, increase, economic, country, unemployment, economy, inflation, increased, time & Macroeconomics\\
60 & care, aged, veterans, community, services, support, home, nursing, child\_care, childcare & Health\\
\addlinespace
61 & bank, commonwealth\_bank, banking, private, credit, money, savings, treasurer, board, trading & Domestic Commerce\\
62 & research, tobacco, scientific, fishing, disease, quarantine, science, fisheries, health, fish & Agriculture\\
63 & law, person, offence, criminal, police, crime, offences, evidence, penalty, attorneygeneral & Law and Crime\\
64 & agreement, trade, local, grants, governments, council, development, financial, national, conference & Foreign Trade\\
65 & democrats, issue, committee, question, time, million, pmi, report, issues, process & Government Operations\\
\addlinespace
66 & war, soldiers, service, country, returned, time, ill, forces, military, soldier & Defense\\
67 & electoral, vote, election, voting, votes, system, party, electors, elections, candidates & Government Operations\\
68 & land, settlement, property, lands, country, lease, leases, pastoral, money, acres & Agriculture\\
69 & information, amendments, support, ensure, report, services, national, review, financial, provide & Government Operations\\
70 & wheat, wool, growers, industry, farmers, board, prices, scheme, marketing, production & Agriculture\\
\addlinespace
71 & industry, export, meat, market, dairy, farmers, producers, levy, wine, rural & Foreign Trade\\
72 & expenditure, loan, increase, revenue, amount, total, budget, money, financial, estimated & Macroeconomics\\
73 & question, time, matter, desire, regard, learned, opinion, deal, position, party & Government Operations\\
74 & budget, tax, billion, million, per\_cent, business, economy, support, jobs, governments & Domestic Commerce\\
75 & millen, senators, question, mcgregor, de, dobson, givens, lt, clemons, time & Government Operations\\
\addlinespace
76 & company, oil, companies, industry, coal, profits, capital, business, private, mining & Energy\\
77 & per\_cent, industry, tax, policy, time, governments, economic, system, program, national & Macroeconomics\\
78 & clause, provision, section, proposed, agreed, committee, words, provisions, person, matter & Government Operations\\
79 & aboriginal, per\_cent, program, governments, commission, question, assistance, funds, development & Social Welfare\\
80 & family, children, families, parents, income, welfare, time, poverty, allowance, parent & Social Welfare\\*
\end{longtable}
\endgroup{}

\begin{figure}
\includegraphics[width=1\linewidth]{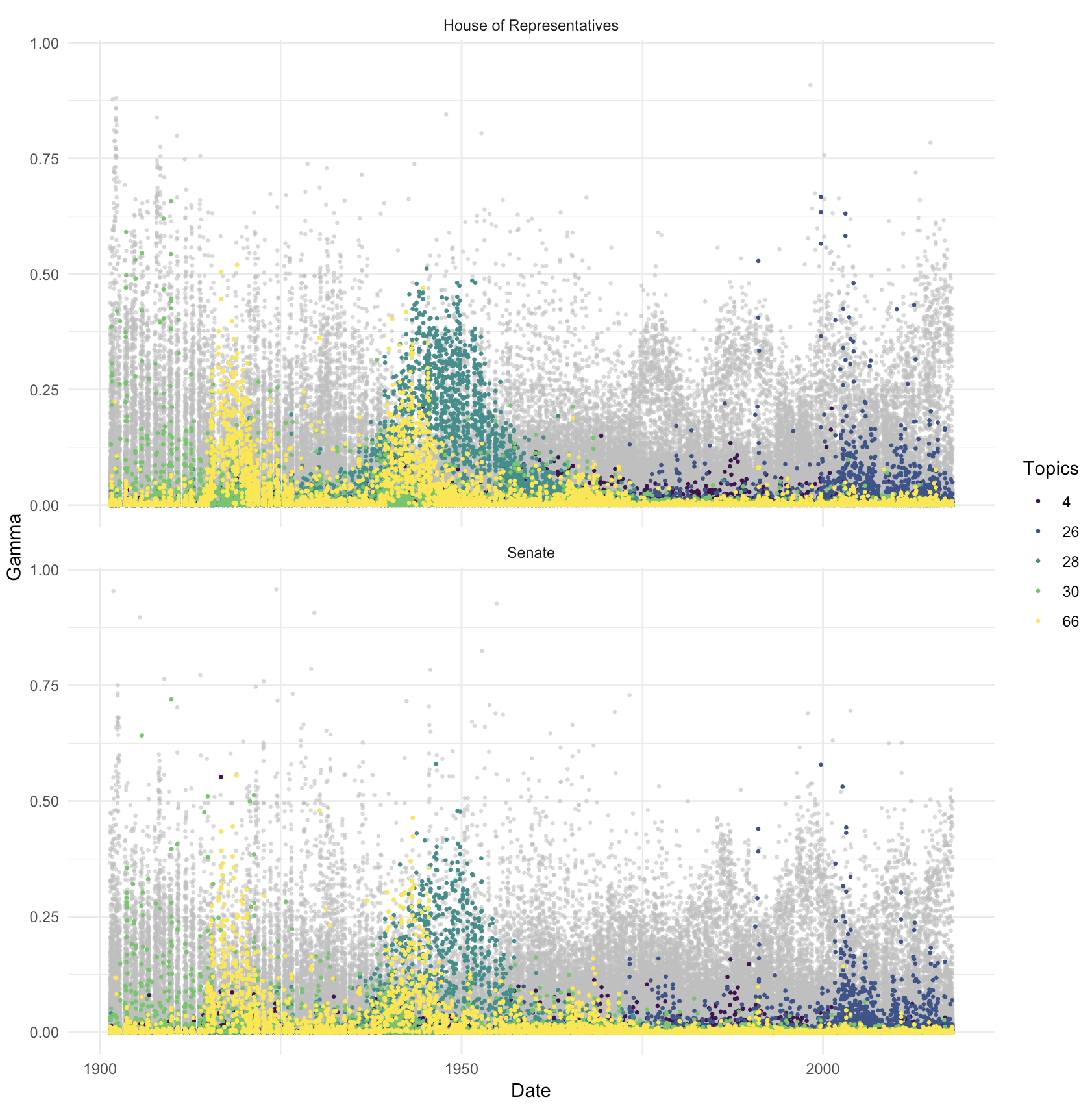} \caption{Illustrative topic model output, with five topics highlighted}\label{fig:exampletopics}
\end{figure}

Figure \ref{fig:exampletopics} illustrates the CTM output based on a sample from the Hansard. The two panels refer to the two different chambers of the Australian Federal Parliament. The highlighted topics are categorised as `Defense' in CAP terms, i.e.~those related to war and conflict. The figure shows how each day's parliamentary discussion can be apportioned to a topic and highlights how these proportions change over time; for example, the highlighted defense topics have notable peaks around the two World Wars, and post-9/11 attacks. The two chambers appear to be similar at a broad scale in terms of the increases and decreases in the different topics. Appendix \ref{topicmodeleconresultsdiscussion} provides another example from the topic model results in the context of economic events.

\begin{figure}
\includegraphics[width=1\linewidth]{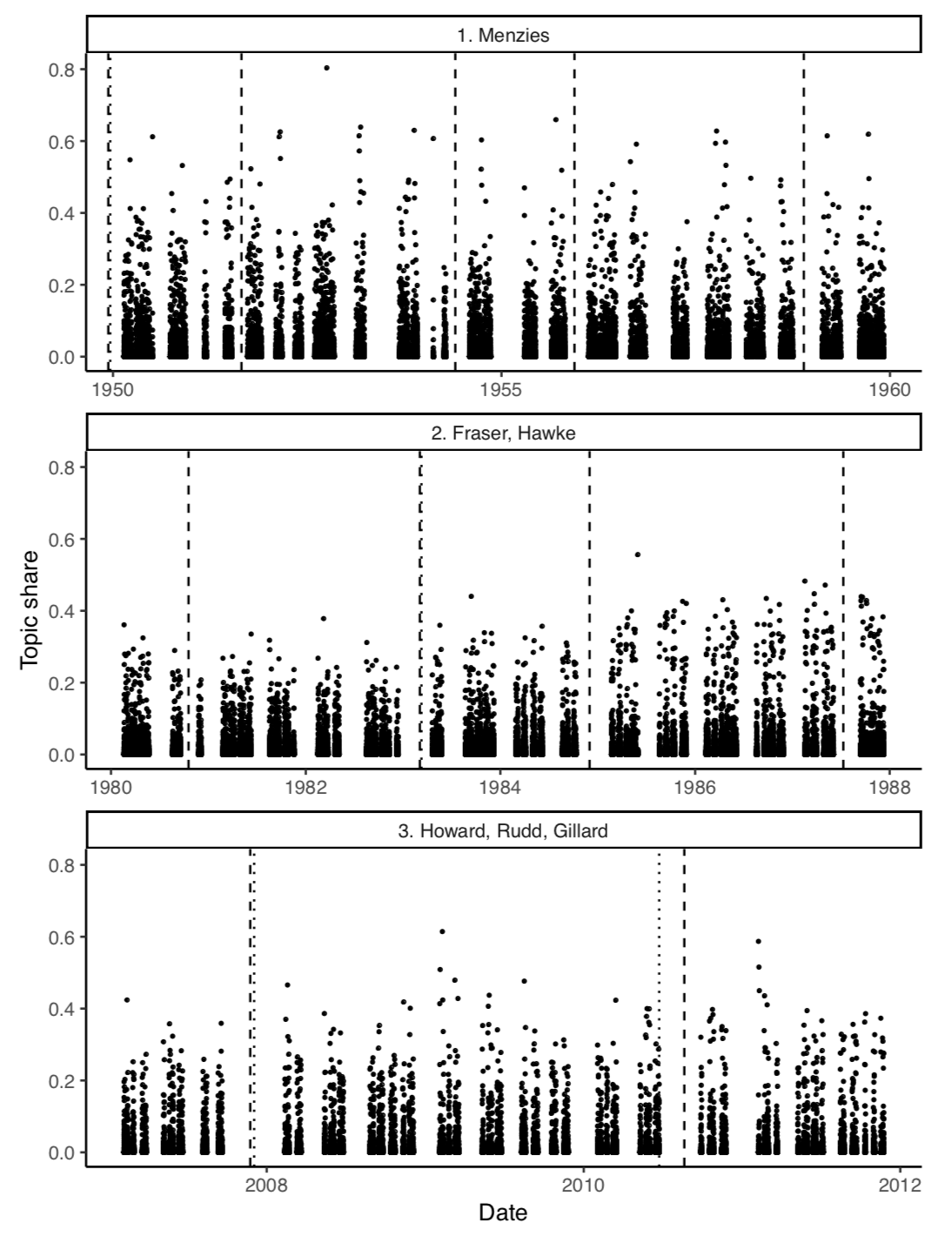} \caption{Illustrative topic model output, for three periods}\label{fig:threeexamples}
\end{figure}

Figure \ref{fig:threeexamples} shows the topic model output in the context of three notable periods of Australian political history.
The first panel of Figure \ref{fig:threeexamples} is the second Menzies term. The dashed lines show the elections. While the results here abstract from differences based on topic mix, the shape of the distribution seems to be reasonably similar. Given 80 topics, it may be difficult to see differences in the topic mix manually, and so we would like for the analysis model to be able to suggest whether the mix is changing. The second panel shows the 1983 election, and the subsequent change from Fraser to Hawke on 11 March 1983, identified by a dotted line on that date. The distribution of topics seems less bunched after the change, but it is difficult to tell how different it is. We would like the analysis model to be able to distinguish between the two if they seem different. The third panel shows the first Rudd term. The important aspect to note is that the period after the 2007 election and the Rudd term contain essentially the same Hansard dates. The analysis model needs to be able to deal with this situation.

\hypertarget{results-from-analysis-model}{%
\subsection{Results from analysis model}\label{results-from-analysis-model}}

As discussed in Section \ref{analysismodelexplanation}, the modelling process takes the topic proportions estimated by the CTM and reduced to the Comparative Agendas Project, and examines the association between these topic distributions and events.

\begin{figure}
\includegraphics[width=1\linewidth]{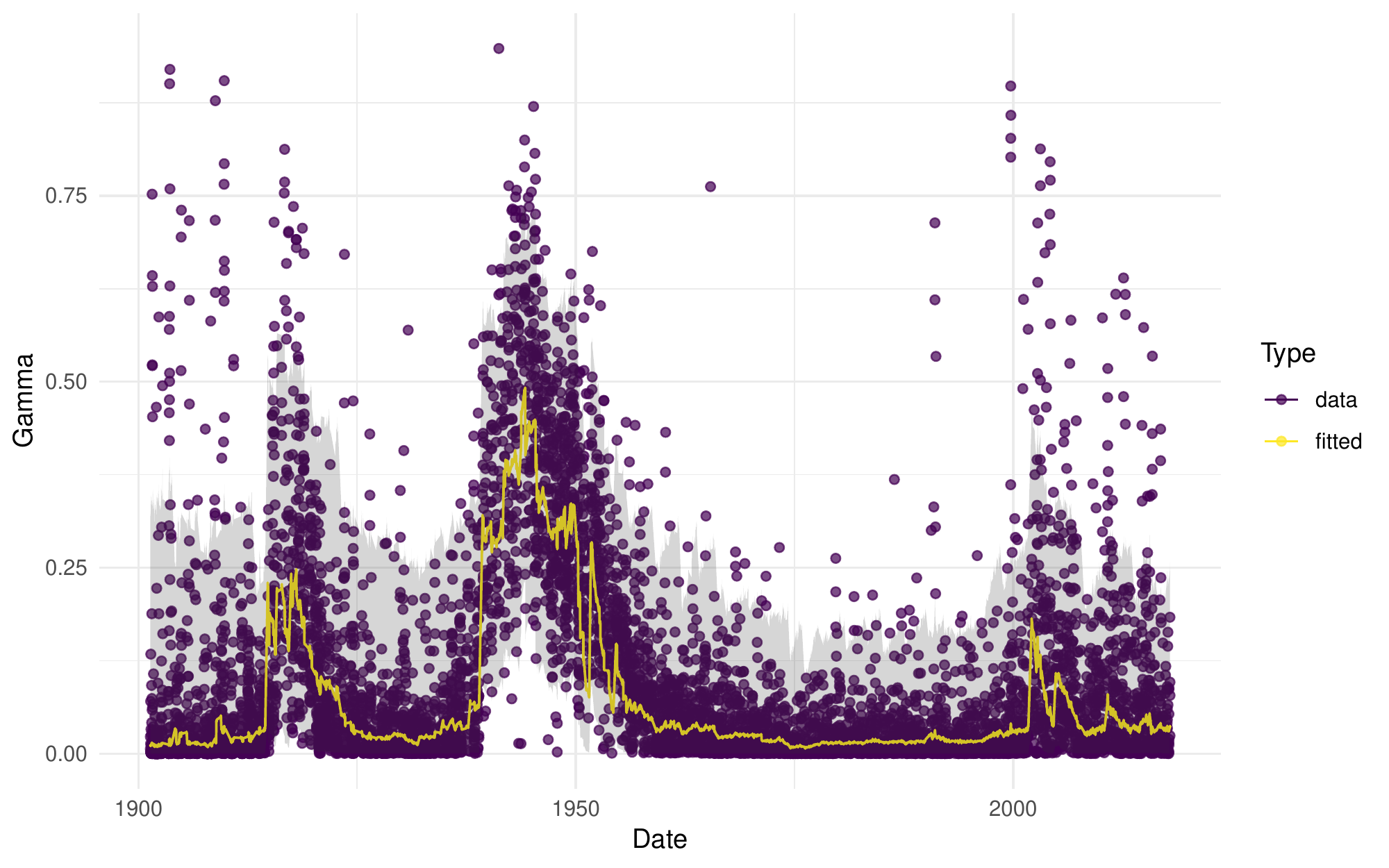} \caption{Model estimates of topic prevalance by sitting period for those related to war and conflict}\label{fig:wartopicsgraph}
\end{figure}

There are several outputs of interest from this modelling stage. For example, the model provides estimates of topic prevalence by each sitting period. Figure \ref{fig:wartopicsgraph} illustrates the estimated topic prevalence by sitting period for topic 15, one of the defense-related topics, overlaid with the daily topic prevalence (as also illustrated in Figure \ref{fig:threeexamples} above). This nicely illustrates how the topics change over time, as the daily estimates tend to be quite variable.

One of the main goals of the analysis model is to see which elections and changes of prime minister are associated with changes in the prevalence of topics over time. By way of background, as Australia has a parliamentary system it is possible for the prime minister to change without an election. If a person was prime minister more than once then these periods are considered independently.

As detailed in Section \ref{analysismodelexplanation}, the model estimates an effect for each prime minister, \(\alpha_g\) and each election, \(\beta_e\). We identify differences between neighbouring prime ministers and between neighbouring elections based on calculating 95 per cent Bayesian credible intervals from posterior samples of these respective mean effects. When these do not overlap we consider that the model finds a difference between either the neighbouring prime ministers or elections, as appropriate.

We summarise our results in terms of prime ministers in Table \ref{tab:primeministersresults} and in terms of elections in Table \ref{tab:electionsresults}. These tables focus on prime ministers and elections that were different to the ones that preceded them.\footnote{Complete lists of the Australian elections and prime ministers are available in Appendices \ref{fulllistofelections} and \ref{fulllistofgovernments}.} A total of 17 out of the 36 prime ministerial periods had significantly different topic discussions than their predecessor. In terms of elections, 12 out of a total of 45 elections were significantly different to the previous cycle.

\begin{table}

\caption{\label{tab:primeministersresults}Prime ministers that were significantly different to their predecessor}
\centering
\fontsize{10}{12}\selectfont
\begin{threeparttable}
\begin{tabular}[t]{rllll}
\toprule
Number & Premiership & Start & End & Party changed\\
\midrule
2 & Deakin 1 & 1903-09-24 & 1904-04-27 & No\\
3 & Watson & 1904-04-27 & 1904-08-18 & Yes\\
6 & Fisher 1 & 1908-11-13 & 1909-06-02 & Yes\\
9 & Cook & 1913-06-24 & 1914-09-17 & Yes\\
12 & Bruce & 1923-02-09 & 1929-10-22 & Yes\\
\addlinespace
14 & Lyons & 1932-01-06 & 1939-04-07 & Yes\\
17 & Fadden & 1941-08-28 & 1941-10-07 & Yes\\
20 & Chifley & 1945-07-13 & 1949-12-19 & No\\
22 & Holt & 1966-01-26 & 1967-12-19 & No\\
24 & Gorton & 1968-01-10 & 1971-03-10 & Yes\\
\addlinespace
25 & McMahon & 1971-03-10 & 1972-12-05 & No\\
28 & Hawke & 1983-03-11 & 1991-12-20 & Yes\\
29 & Keating & 1991-12-20 & 1996-03-11 & No\\
31 & Rudd 1 & 2007-12-03 & 2010-06-24 & Yes\\
32 & Gillard & 2010-06-24 & 2013-06-27 & No\\
\addlinespace
34 & Abbott & 2013-09-18 & 2015-09-15 & Yes\\
35 & Turnbull & 2015-09-15 & 2018-08-24 & No\\
\bottomrule
\end{tabular}
\begin{tablenotes}[para]
\item \textit{Note: } 
\item The significance of a prime minister is determined by whether at least one topic was significantly different during this term, compared with the previous one.
\end{tablenotes}
\end{threeparttable}
\end{table}

\begin{table}

\caption{\label{tab:electionsresults}Election-periods that were significantly different to the one before}
\centering
\fontsize{10}{12}\selectfont
\begin{threeparttable}
\begin{tabular}[t]{rrlrll}
\toprule
Number & Year & Date & Total seats & Election winner & Changed party\\
\midrule
9 & 1922 & 1922-12-16 & 75 & Non-labor & No\\
19 & 1949 & 1949-12-10 & 121 & Non-labor & Yes\\
20 & 1951 & 1951-08-28 & 121 & Non-labor & No\\
22 & 1955 & 1955-12-10 & 122 & Non-labor & No\\
24 & 1961 & 1961-12-09 & 122 & Non-labor & No\\
\addlinespace
25 & 1963 & 1963-11-30 & 122 & Non-labor & No\\
29 & 1974 & 1974-05-18 & 127 & Labor & No\\
33 & 1983 & 1983-03-05 & 125 & Labor & Yes\\
34 & 1984 & 1984-12-01 & 148 & Labor & No\\
38 & 1996 & 1996-03-02 & 148 & Non-labor & Yes\\
\addlinespace
43 & 2010 & 2010-08-21 & 150 & Labor & No\\
45 & 2016 & 2016-07-02 & 150 & Non-labor & No\\
\bottomrule
\end{tabular}
\begin{tablenotes}[para]
\item \textit{Note: } 
\item The significance of an election is determined by whether at least one topic was significantly different during the period between this election and the next, compared with the period between the previous election and this one.
\end{tablenotes}
\end{threeparttable}
\end{table}

\begin{figure}
\includegraphics[width=0.9\linewidth]{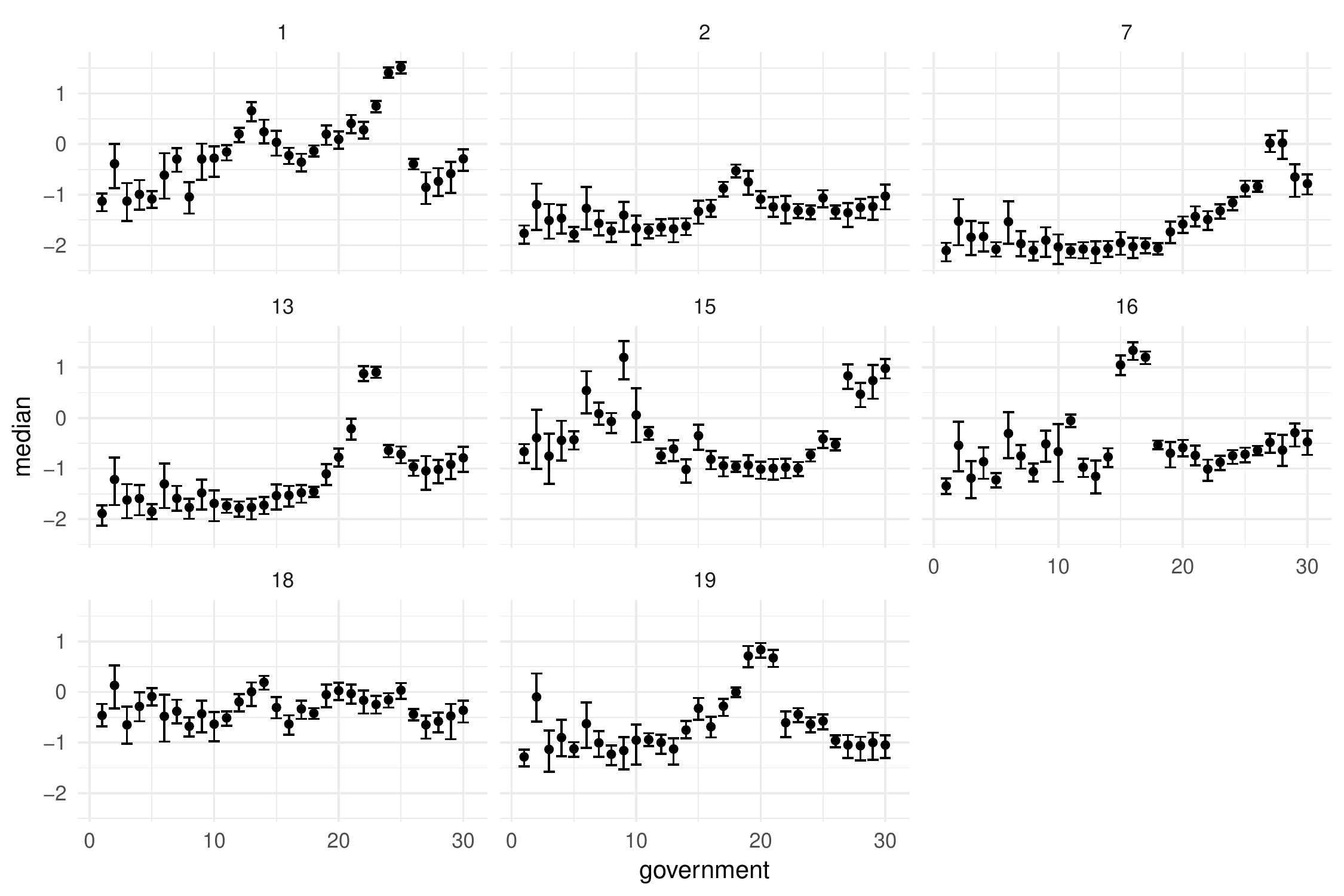} \caption{Level effects for prime ministers by selected topics.}\label{fig:mugov}
\end{figure}

\begin{figure}
\includegraphics[width=0.9\linewidth]{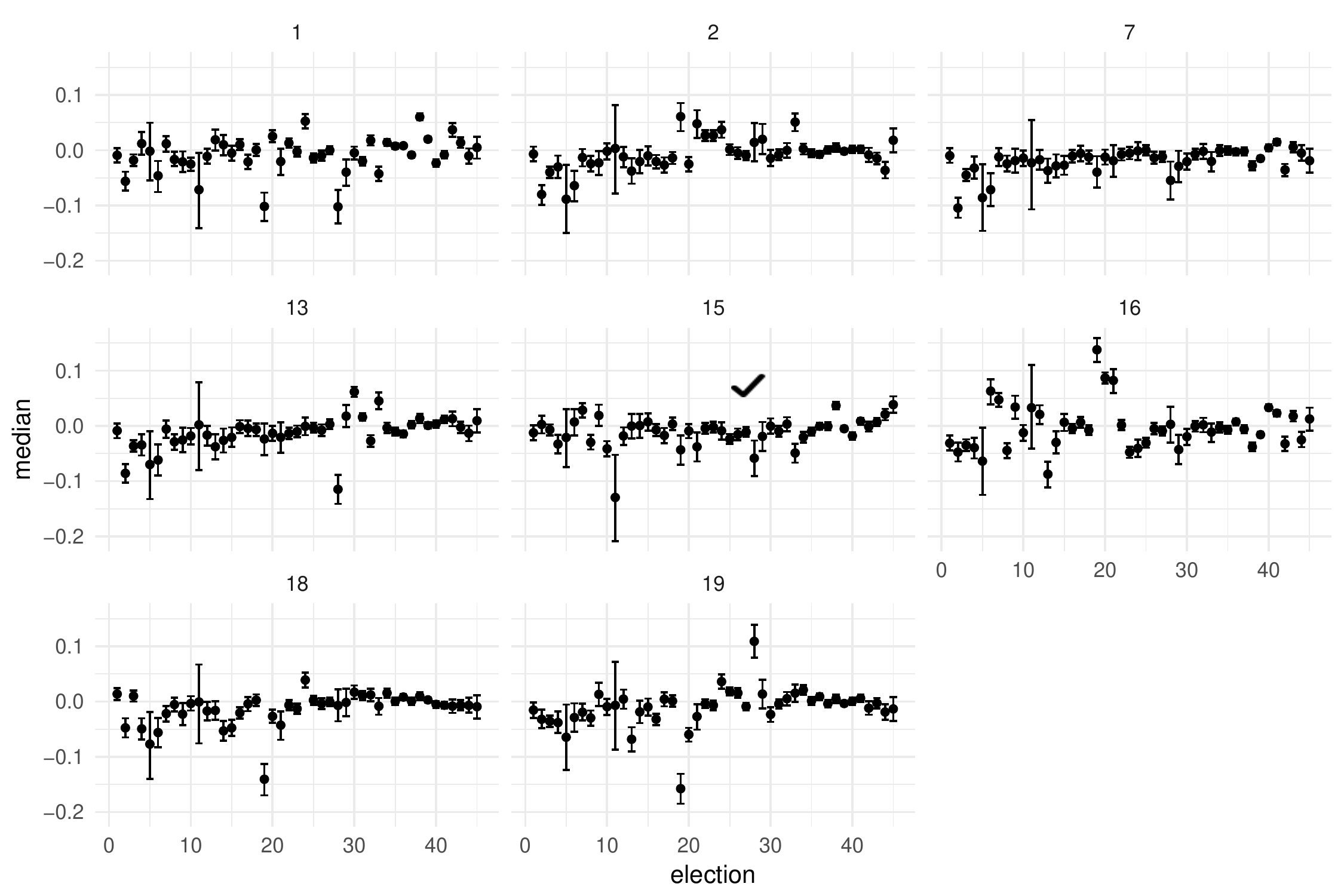} \caption{Level effects for election periods by selected topics}\label{fig:muelection}
\end{figure}

In Figures \ref{fig:mugov} and \ref{fig:muelection} we focus on certain topics to illustrate significant differences between prime ministers and elections, respectively. In the graphs, the points show the estimated value of \(\alpha_g\) and \(\beta_e\), respectively, for each of the topics specified. The error bars represent 95 per cent Bayesian credible intervals. We have focused on Topics 1, 2, 7, 13, 15, 16, 17, and 19 here which have to do with war and conflict. Large differences can be seen at various times.

\hypertarget{additional-results-identifying-outlying-days}{%
\subsubsection{Additional results: identifying outlying days}\label{additional-results-identifying-outlying-days}}

Once prime minister and election effects have been taken into consideration, some days stand out compared with others in their sitting period. We do not explicitly include them in the model because of over-fitting and effect-type concerns, but we are interested to see if these can be explained by events that occurred on or before that sitting day. For instance, we may expect that events of a historical magnitude, such as the 9/11 attacks would change the discussion, or that the sitting day when, for instance, the Apology to the Stolen Generation was delivered, or some particularly prominent legislation introduced, would be different to others in that sitting period.

The model set-up allows for these outlying days to be identified. In particular, we are estimating \(\mu_{c,s,p}\), which can be thought of as the mean topic prevalence for topic \(p\) in chamber \(c\) and sitting period \(s\). We can identify `outlying days' by comparing this mean prevalence and a particular day's topic distribution. This approach means that the model generates dates that are interesting without us having to specify interesting dates. More specifically, we define a day to be `outlying' or `different-to-expected' if the topic distribution on that particular day is more than three standard deviations different to the mean topic distribution for the relevant sitting period.

Table \ref{tab:keyeventsresultstable} summarises the days where parliamentary discussion was significantly different from the rest prevailing in that week. Notably, considerably more than a majority of these dates occur in the first half of our sample. And a full seven occur in 1909.

\begin{table}

\caption{\label{tab:keyeventsresultstable}Days that were significantly different to other days in their sitting period}
\centering
\fontsize{10}{12}\selectfont
\begin{threeparttable}
\begin{tabular}[t]{llllll}
\toprule
Dates &  &  &  &  & \\
\midrule
1904-07-15 & 1911-12-13 & 1930-06-06 & 1945-05-30 & 1991-01-22 & 2002-08-26\\
1908-10-30 & 1912-10-02 & 1930-06-11 & 1945-06-12 & 1999-09-21 & 2003-03-20\\
1909-10-13 & 1912-10-18 & 1930-06-12 & 1946-11-06 & 1999-09-22 & 2003-05-13\\
1909-10-14 & 1923-07-27 & 1935-10-09 & 1947-11-14 & 1999-09-23 & 2006-02-16\\
1909-10-19 & 1923-07-30 & 1935-11-06 & 1949-02-22 & 1999-09-27 & 2010-03-16\\
\addlinespace
1909-10-20 & 1923-07-31 & 1935-11-07 & 1954-02-15 & 2001-03-08 & 2012-10-30\\
1909-10-21 & 1923-08-01 & 1944-03-17 & 1959-03-12 & 2001-05-09 & 2013-06-25\\
1909-10-22 & 1925-08-26 & 1944-07-17 & 1965-05-26 & 2001-09-17 & 2015-11-30\\
1909-10-26 & 1927-10-19 & 1945-04-18 & 1991-01-21 & 2002-06-04 & -\\
\bottomrule
\end{tabular}
\begin{tablenotes}[para]
\item \textit{Note: } 
\item These days were significantly different to others in their sitting period after taking election and prime minister effects into consideration.
\end{tablenotes}
\end{threeparttable}
\end{table}

\hypertarget{discussion}{%
\section{Discussion}\label{discussion}}

In this paper we explored changes in parliamentary discussion in association with changing prime minister and elections events in Australia. We used OCR and other text processing techniques to create a comprehensive dataset which covers parliamentary discussion in Australia throughout history since Federation in 1901 to the present day. We then used topic modeling to ascertain comparable topics of discussion over time, and a Bayesian hierarchical Dirichlet model to relate these topics to changes in elections and prime ministers. In general we find that changes in prime minister change the distribution of topics discussed in parliament, but that most elections did not. We find that significant events such as 9/11 and the Bali Bombings had a substantial effect.

Of the 36 prime ministers over this period, we find that 17 of them were significantly different to the prime minister that preceded them, after we account for and remove some that had especially short terms. The three earliest of them occur in the first decade after Federation, and to a certain extent this may be due to the variety of the issues that had to be addressed in those initial years. Joseph Cook's time as prime minister is found to be significantly different to that of his predecessor, Andrew Fisher, although it may be that this is due to World War I.

The second Menzies term, beginning in 1949, is the next government that is significantly different to its predecessor. The other governments that are different are concentrated in the second half of our sample, with three of them being in the past twenty years. Similarly, of the 45 general elections that have been held we find that 14 of them define periods that were significantly different to their immediate predecessor. 1974, 1980, 1990, 1998, 2004, and 2007 stand out as elections where the government did not change, but the model suggests there was considerable change in the topics discussed in parliament.

The second Menzies term was associated with a variety of changes compared with the preceding Chifley Government. Chifley had governed through the end of World War II and the difficult economic times that followed. There was also a large increase in the number of seats in the House of Representatives at the 1949 election. Many new politicians entered parliament, and this changed representation may also have been partly to do with the changed distribution of discussion topics, although further investigation of this is left for future work. The sixteen-year length of the second Menzies term, and better economic conditions over this time make it understandable why parliamentary discussion would have been different. There were six elections within the second Menzies term. Three toward the middle of that government were associated with significant changes in the topics discussed.

Menzies was succeeded by Holt in January 1966. This is an example where there was a change in prime minister without an election, as the next election only happened in November 1966. We find that Holt is different to Menzies.

\begin{figure}
\includegraphics[width=1\linewidth]{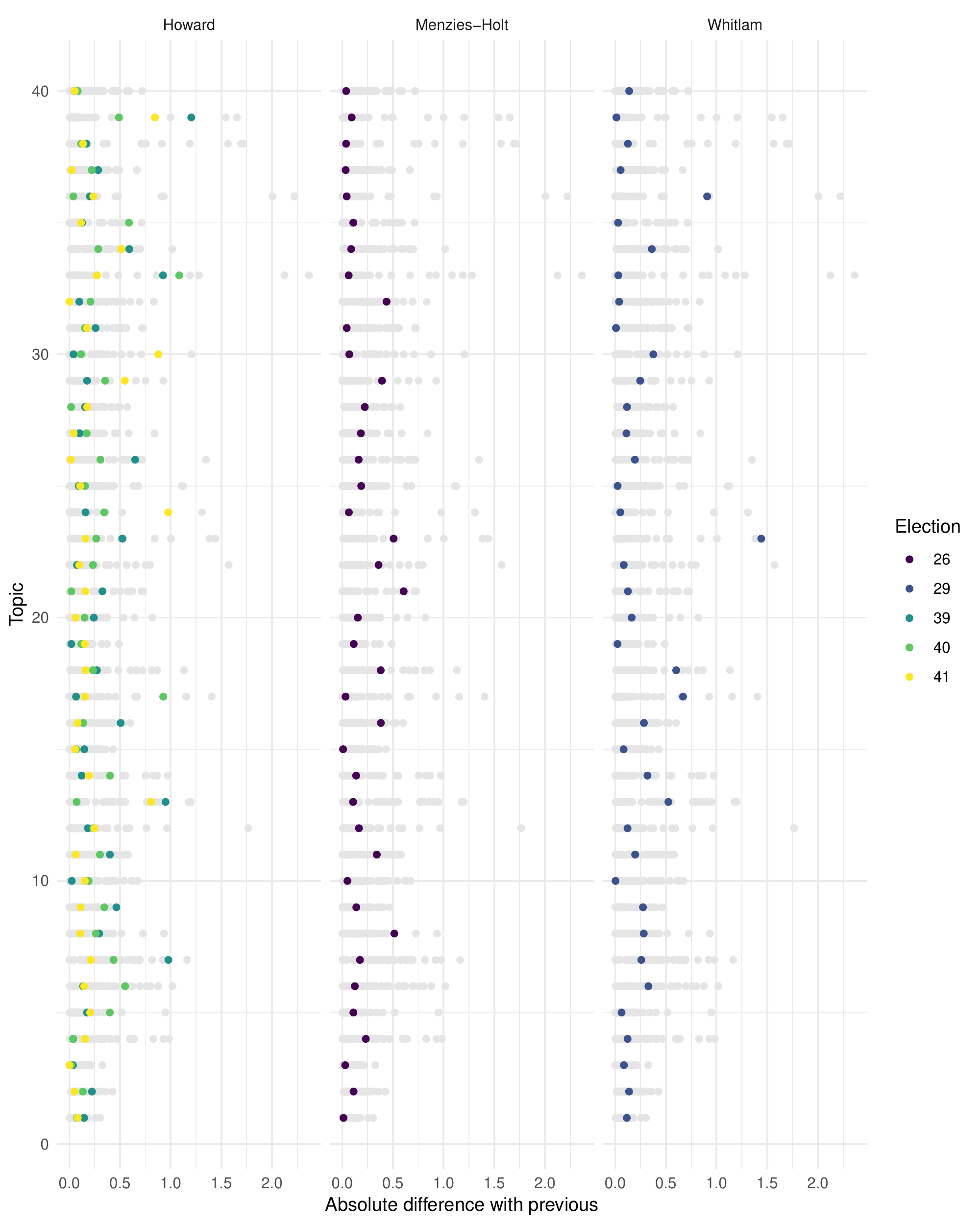} \caption{Differences between various elections}\label{fig:muelectionall}
\end{figure}

In Figures \ref{fig:muelectionall} we compare the topics during Menzies' final term with the topics of Holt.

Whitlam's term is especially interesting as we find a difference in the topics after he was first elected in December 1972, compared with his second election win in May 1974. Figure \ref{fig:muelectionall} compares the topics that are significant before and after the election.

Howard is also interesting because of the significant differences between elections. For instance, each of the election periods is associated with fairly substantial differences compared with the preceding election periods, and all are actually significantly different at the 95 per cent level. Figure \ref{fig:muelectionall} compares the topics that are significant in the different elections that Howard won.

To a certain extent the change after the November 2001 election is expected because of the 9/11 terrorist attacks that had only occurred two months earlier, the Bali Bombings that occurred in October 2002, and the dramatic increase in the discussion related to terrorism and conflict over these years. However, the change in 1998 and 2004 is more unexpected. Although Howard is the second-longest serving prime minister and commonly thought of as a period of stability because the senior ministers were consistent as well, it might be that it is better to think of Howard as a combination of three or four different periods and that Howard reinvented itself over this period.

One advantage of our analysis model compared with using the STM approach is that we can create a measure that is equivalent to testing for outliers in a model where the underlying variables were not latent. The results of this reduction in supervision are promising, but suggest the specifics of our process may need further refinement. For instance, our approach appropriately identifies the sitting day that first follows 11 September 2001. But there are many dates that we would have expected to be identified that were not, and similarly some of the dates that were identified are surprising. When we examine this we find that some of them are associated with significant legislation. Although our results are not overly over-weighted in the first half of our sample, there is a substantial gap between 1965 and 1991. Table \ref{tab:keyeventsresultstable} highlights further work is needed to improve this approach. For instance, it may be that our approach is not appropriately considering step-changes or it may be that our identification of sitting periods is not appropriate for the entire period, given the changed sitting pattern that we identify.

\hypertarget{limitations-and-future-work}{%
\subsection{Limitations and future work}\label{limitations-and-future-work}}

Using text as data allows us to conduct larger-scale analysis that would not be viable using less-automated approaches and so researchers may be able to identify associations and patterns that would otherwise have been overlooked. That said, the approach has well-known shortcomings and weaknesses, such as those documented by \citet{GrimmerStewart2013}. Because of these, our paper should be considered a complement to more detailed analysis such as qualitative methods and case studies.

One weakness of topic models is that the number of topics needs to be specified but there is rarely a clear reason to choose some particular number. There are also more nuanced weaknesses to be aware of. For instance, the topics need to be interpreted. Some of our topics were not overly meaningful, especially on their own. Interpretation is especially difficult when the number of topics is large, but we found that a large number of topics worked best for our dataset.

Although topic modelling is an unsupervised machine learning technique, the inputs require fine-tuning. For instance, selecting stop words for removal and which words to merge because of common co-location has an impact on the topics. Topic model outcomes are known to be sensitive to preprocessing \citep{DennySpirling2018} and ours was no exception.

One way to get around using topic models is to use a supervised learning approach, such as that used by \citet{AshMorelliOsnabru2018} in the context of New Zealand Hansard. Another is to use the words more directly, for instance word2vec and other approaches. As computational power become cheaper and more appropriate analytical methods, such as \citet{Taddy2015} as applied in \citet{GentzkowShapiroTaddy2018} in the context of examining congressional speech records, are developed this becomes a more feasible options and future research could explore that direction.

In this paper we think of events as affecting daily discussion in parliament. Given the events that we consider and the broad topics of discussion we consider this the most appropriate direction for our paper. However, conducting a similar analysis at a person, instead of daily, level would lead to interesting results. This would allow the analysis model to include a rich set of person-level covariates such as gender or party, and account for broader factors such as the televising of question time, or the state of the economy and the budget position. It would also be especially interesting to consider reversing the direction of causality and examine the effect of what is said in parliament on various economic, political and social events.

In terms of the analysis model that relates the topic distributions with events, there are several limitations to the model. Firstly, we are assuming that the effect of a particular prime minister is constant across the whole period. In addition, we assume that the effect of elections is monotonically decreasing across days since election. Future work could consider different functional forms on both of these effects, and in particular try to allow for elections to have a `lead-up' effect.

The way that we identify unusual periods could also be improved. We defined sitting periods in a constant fashion across the whole dataset time frame, but how long the stretches are that parliament sits for has changed over time. In addition, more work needs to be done on how to identify outlying events. For instance, the extent to which an important event that occurs outside a sitting period can be identified has a great deal of uncertainty. And if an event happens in the middle of a sitting period, it may have a large effect on the overall mean, such that specific days are not identified as significantly outlying.

Finally, the current modelling and analysis set-up is a two-stage process: we take the output of a topic model, and use this as the input to a second model. However, this approach does not appropriately propagate the uncertainty of the topic distribution estimation stage. Future methodological work could consider how to combine these two modelling steps, for instance by extending the STM approach into a more flexible framework.

Watching Australian politicians at work can sometimes be a little disheartening. It can be hard to believe that not only are those in charge shouting insults that would not be tolerated in a schoolyard, but that the electorate voted to put them there. Nonetheless, our work suggests that important topics are discussed in parliament. It is easy to look back and think that we live in uniquely tumultuous times, but our analysis suggests events have always driven debate and that periods of stability may be the exception. However, we do find that since the 1990s the effect of prime ministers and elections on the topics discussed in the Australian parliament does seem more pronounced than it used to be.

\newpage

\hypertarget{appendix-appendix}{%
\appendix}

\textbf{Appendix}

\textbf{Authors:} Rohan Alexander and Monica Alexander

\hypertarget{hansard-details}{%
\section{Hansard details}\label{hansard-details}}

\hypertarget{examplehansardpage}{%
\subsection{Example Hansard page}\label{examplehansardpage}}

\begin{figure}

{\centering \includegraphics[width=0.65\linewidth]{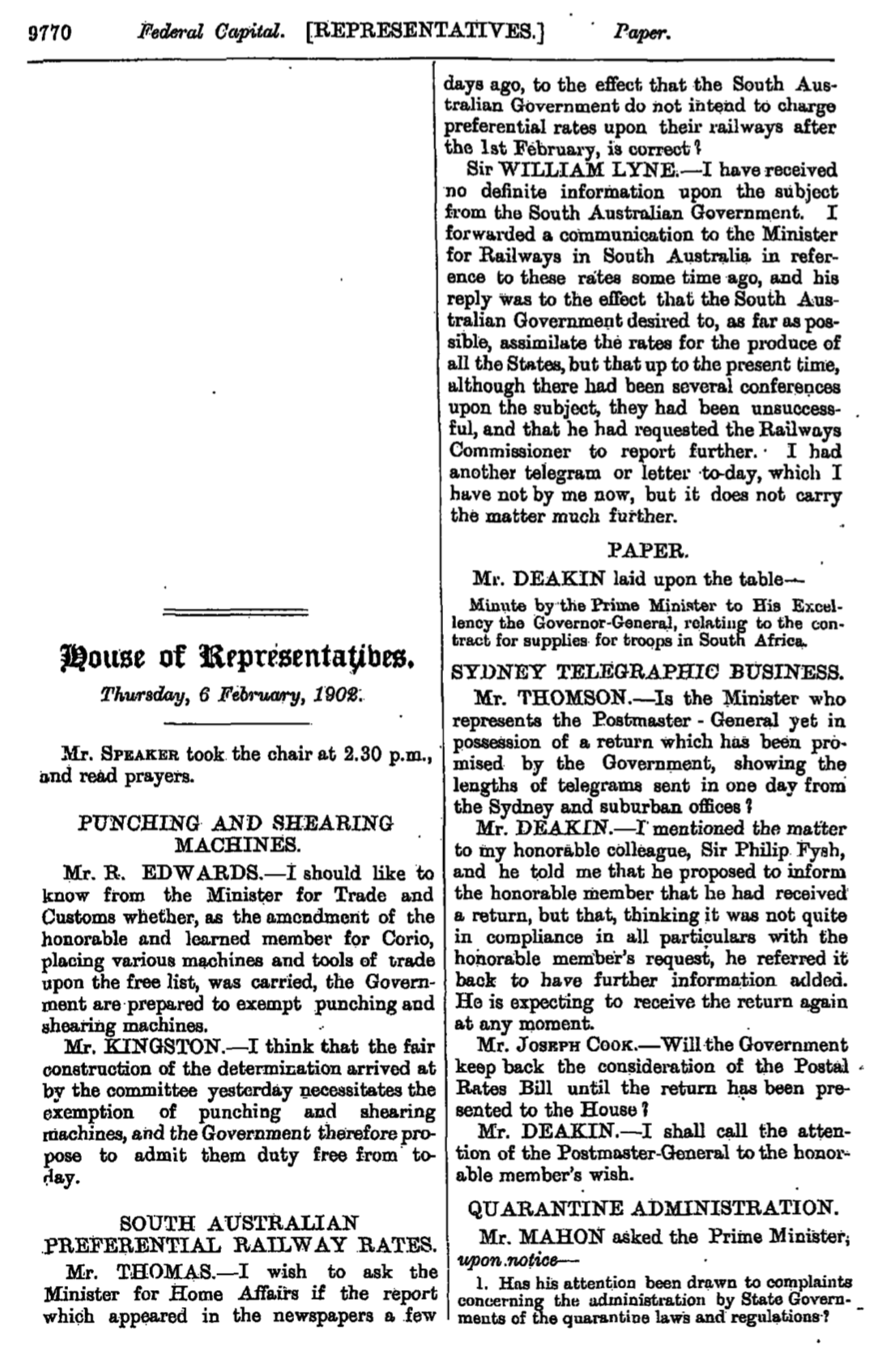} 

}

\caption{Example Hansard page -- 6 February 1902}\label{fig:hansardexamplepage}
\end{figure}

\newpage

\hypertarget{hansardsummarystatistics}{%
\subsection{Summary statistics}\label{hansardsummarystatistics}}

\hypertarget{counts-per-year}{%
\subsubsection{Counts per year}\label{counts-per-year}}

\begin{figure}
\includegraphics[width=1\linewidth]{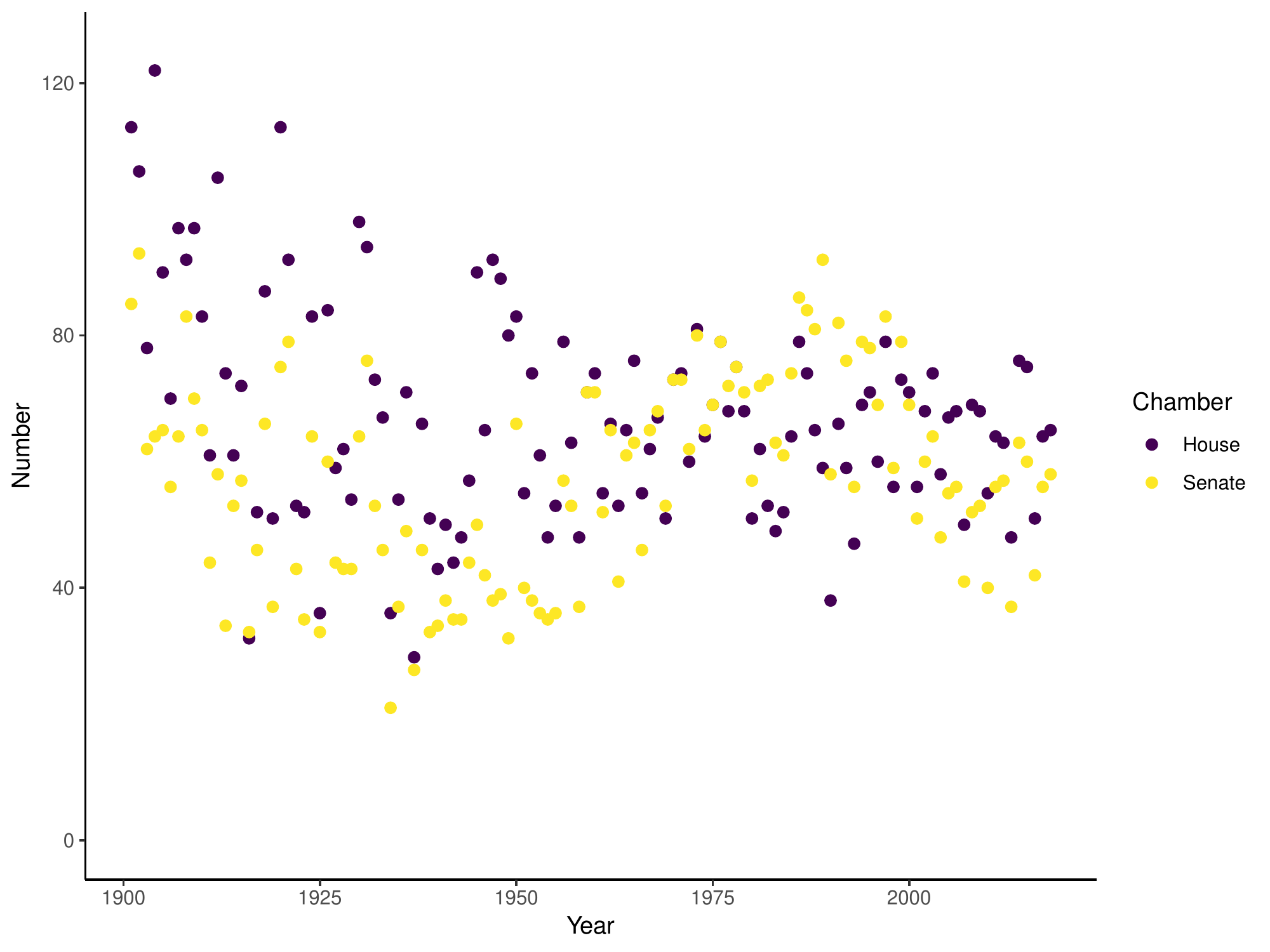} \caption{Number of sitting days, by year}\label{fig:countsbyyear}
\end{figure}

The number of sitting days in a year varies considerably. The highest in the House of Representatives was 122 days in 1904, followed by 113 days in 1901 and 1920. The year with the most sitting days in the Senate was 1902 with 93 days, followed by 1989 with 92 days, and 1986 with 86 days (Figure \ref{fig:countsbyyear}).

Until the 1950s the House of Representatives tended to have more sitting days than the Senate. It was then similar, before the Senate had more days in the 1980s and 1990s. Since the 2000s the House of Representatives again has tended to have more sitting days than the Senate.

\hypertarget{distribution-over-the-year}{%
\subsubsection{Distribution over the year}\label{distribution-over-the-year}}

\begin{figure}
\includegraphics[width=1\linewidth]{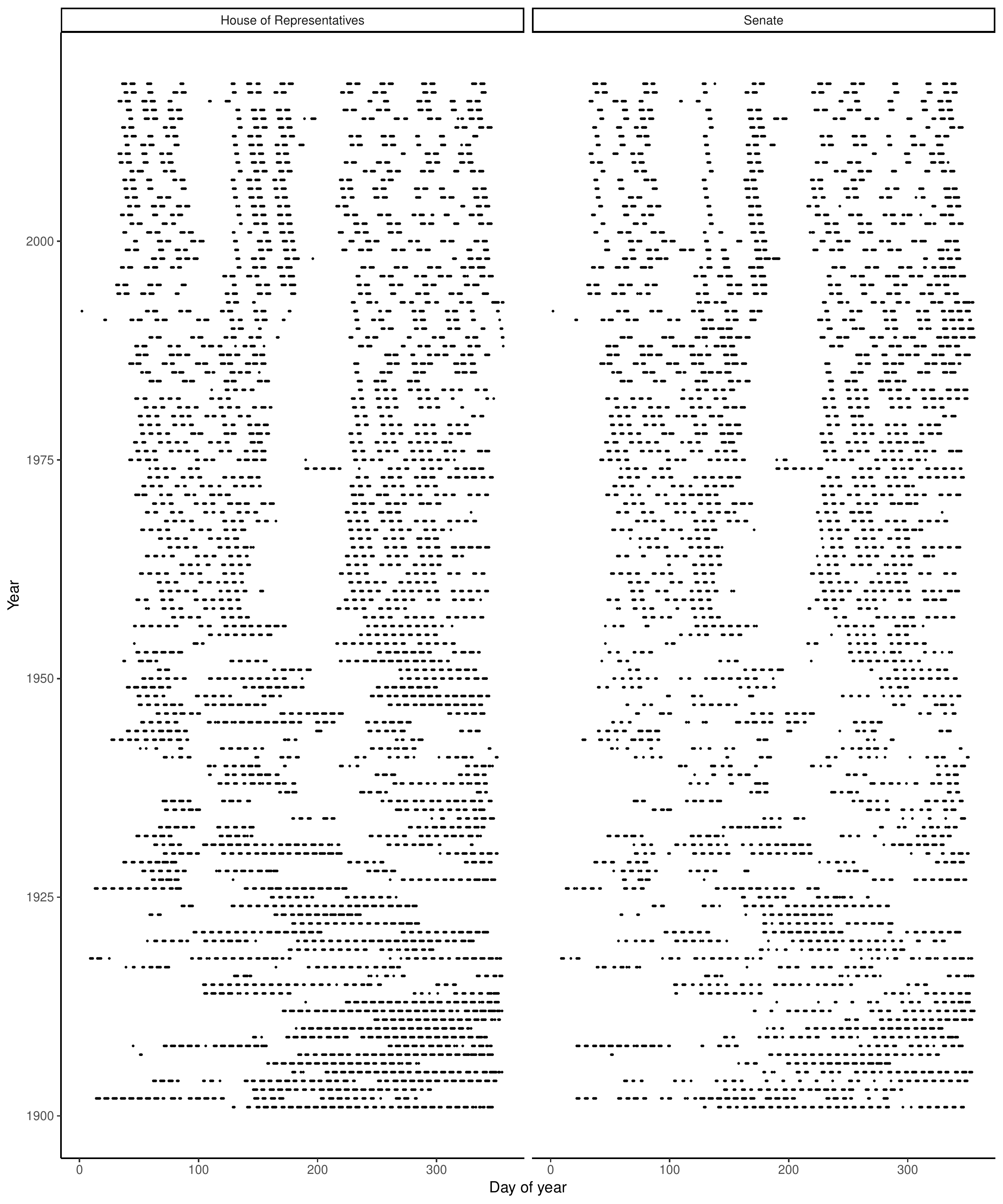} \caption{All sitting days}\label{fig:allsittingdaysplotted}
\end{figure}

The distribution of sitting days over the course of the year changes over time (Figure \ref{fig:allsittingdaysplotted}).

\begin{figure}
\includegraphics[width=1\linewidth]{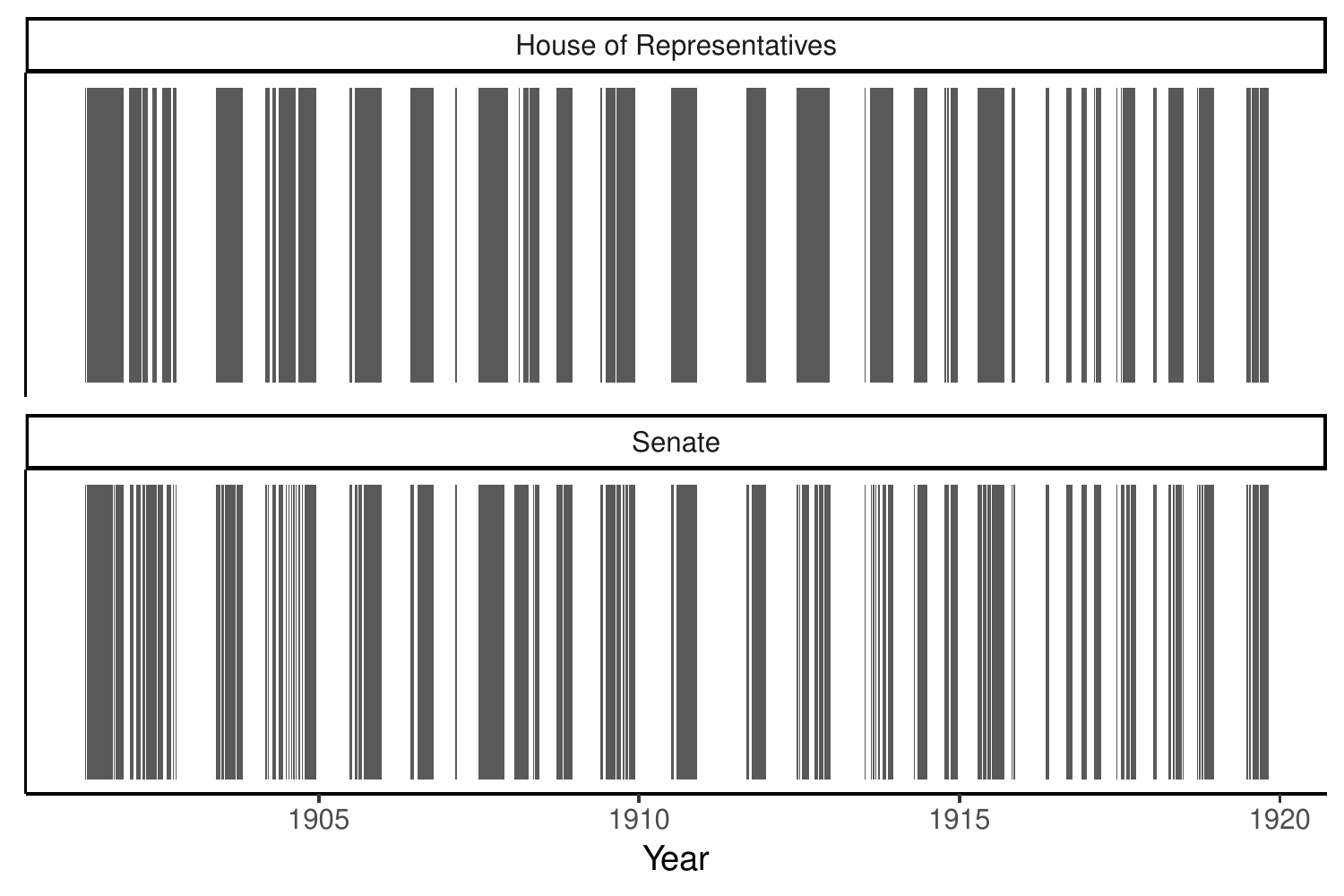} \caption{Number of sitting days, by year}\label{fig:dayssattonineteentwenty}
\end{figure}

\begin{figure}
\includegraphics[width=1\linewidth]{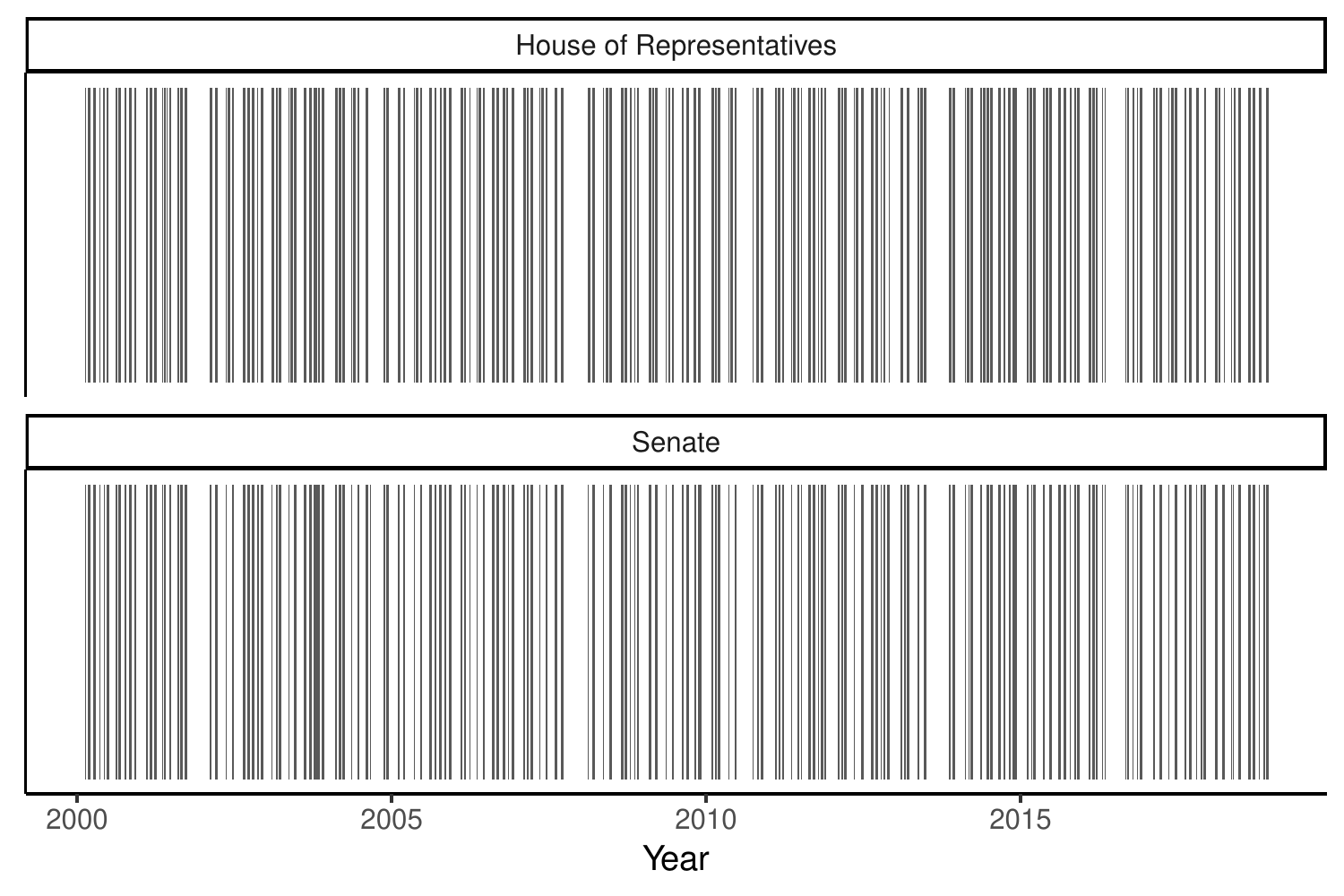} \caption{Number of sitting days, by year}\label{fig:dayssatfromtwothousand}
\end{figure}

It was initially more piecemeal. This can be seen by comparing the pattern of sitting days in the years to 1920 (Figure \ref{fig:dayssattonineteentwenty}) with those in the 19 years from, and including, 2000 (Figure \ref{fig:dayssatfromtwothousand}).

The largest gap between sitting dates for both the House of Representatives and the Senate is 284 days, which happened when neither house sat between 25 November 1910 and 5 September 1911. The next longest gap is 244 the House of Representatives and 243 for the Senate when the lower house last sat on 9 October 1924, the upper house last sat on 10 October 1924 and neither sat again until 10 June 1925.

These counts of the number of sitting days are based on available PDFs. For this reason the counts may be slightly different to other counts. An example of one known issue of this type is detailed in the next section.

\newpage

\hypertarget{knownhansardissues}{%
\subsection{Annual counts of sitting days, compared with parliamentary website}\label{knownhansardissues}}

The parliamentary website provides a summary table of the number of sitting days in each year by chamber.\footnote{As at 5 November 2018, the website was available at: \url{https://www.aph.gov.au/Parliamentary_Business/Statistics/Senate_StatsNet/General/sittingdaysyear}.} Comparing the numbers provided in that table with number of days that we have provides an indication of how complete our dataset is.

\begin{figure}
\includegraphics[width=1\linewidth]{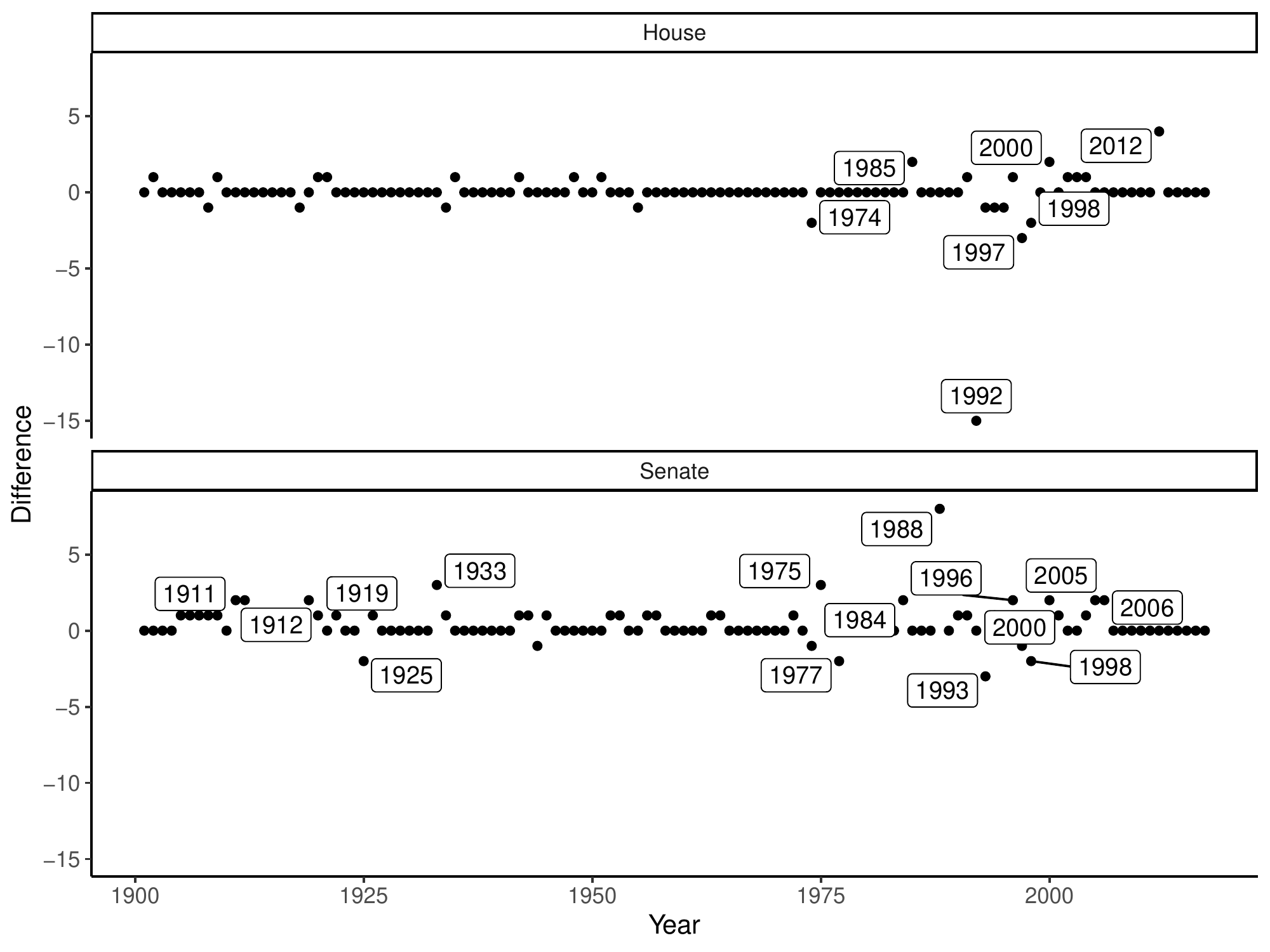} \caption{Differences by year between the number of sitting days and our number of PDFs}\label{fig:differencesbyyear}
\end{figure}

In general the number of sitting days on the parliamentary website summary table is similar to the number of PDFs that we have although it does identify a few particularly concerning years (Figure \ref{fig:differencesbyyear}).

When the difference is positive, it means that in that year we have fewer PDFs than the parliamentary website claims. For instance, 5 could mean that the parliamentary website claimed there were 100 days, but we only had PDFs for 95 days. Similarly, when the difference is negative then we have more PDFs than the parliamentary website claims there were sitting days.

The two major years of concern are 1992 in the House of Representatives where we have 15 days more than the parliamentary website claims there were, and 1988 in the Senate where we have eight days fewer. We examined the physical copy of the Hansard kept in the NSW State Library and this suggests that the summary table on the parliamentary website may be wrong.

The Parliament website is missing the Hansard PDFs for the following dates in the Senate:
1988-12-21,
1988-12-20,
1988-12-19,
1988-12-16,
1988-12-15,
1988-12-14,
1988-12-13,
1988-12-12,
2000-10-12,
2000-06-19, and
2004-08-09.

There are two unaccounted for differences in 2006, one unaccounted for difference in 2001.

The Parliament website is missing the Hansard PDFs for the following dates in the House of Representatives:
1985-08-23,
1992-09-10,
1996-12-13,
2000-10-12,
2000-06-29, and
2002-05-14.

There is one unaccounted for difference in 1920, 1921, 1935, 1942, 1948, 1951, 1991, 2003, 2004, and there are two unaccounted for in 1985 and four unaccounted for in 2012.

In terms of other known issues, in the Senate, the PDF for the website date 10 August 1917 may be wrong. When downloaded the PDF says that it is for 10 January 1918 on the cover sheet, but there's no website entry for 10 January 1918. This is also the case for 18 December 1918 (which contains the PDF for 28 November 1918), and for 1 August 1917 (which contains the PDF for 10 August 1917).

\newpage

\hypertarget{stopwordsgraph}{%
\subsection{Stopwords over time}\label{stopwordsgraph}}

\begin{figure}
\includegraphics[width=1\linewidth]{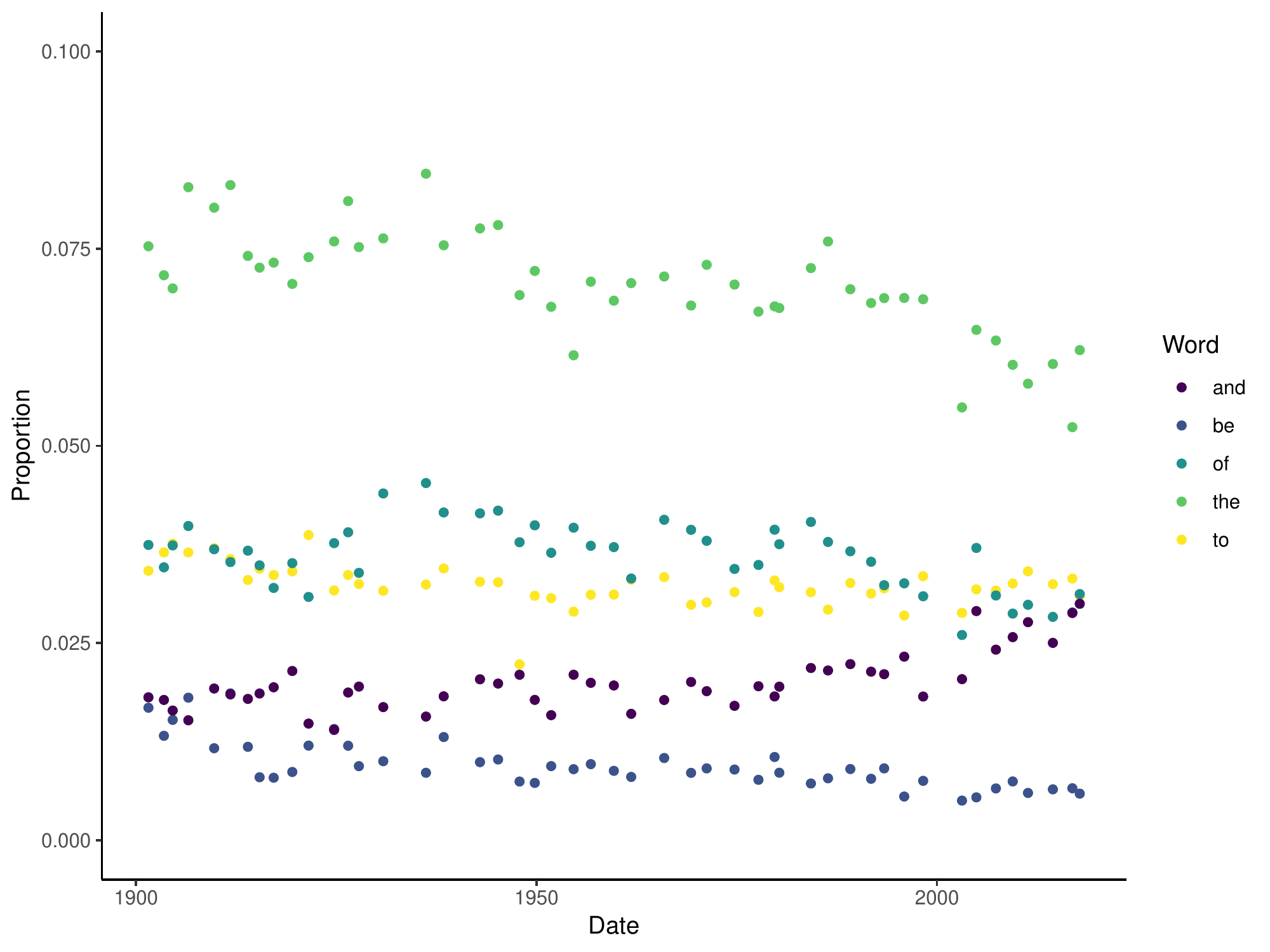} \caption{Proportion that some common words comprise of all words, over time}\label{fig:stopwordsproportion}
\end{figure}

Figure \ref{fig:stopwordsproportion} shows the proportion of five common words -- `and', `be', `of', `the', `to' -- compared with the total number of words over time.

\newpage

\hypertarget{LDAexample}{%
\section{Topic modelling example and details}\label{LDAexample}}

\hypertarget{LDAoverviewandexample}{%
\subsection{Overview and example}\label{LDAoverviewandexample}}

As applied to Hansard, LDA considers each statement to be a result of a process where a politician first chooses the topics they want to speak about. After choosing the topics, the politician then chooses appropriate words to use for each of those topics. Statistically, LDA considers each document as having been generated by some probability distribution over topics. Similarly, each topic is considered a probability distribution over terms. To choose the terms used in each document, terms are picked from each topic in the appropriate proportion.

\begin{figure}
\centering
\includegraphics{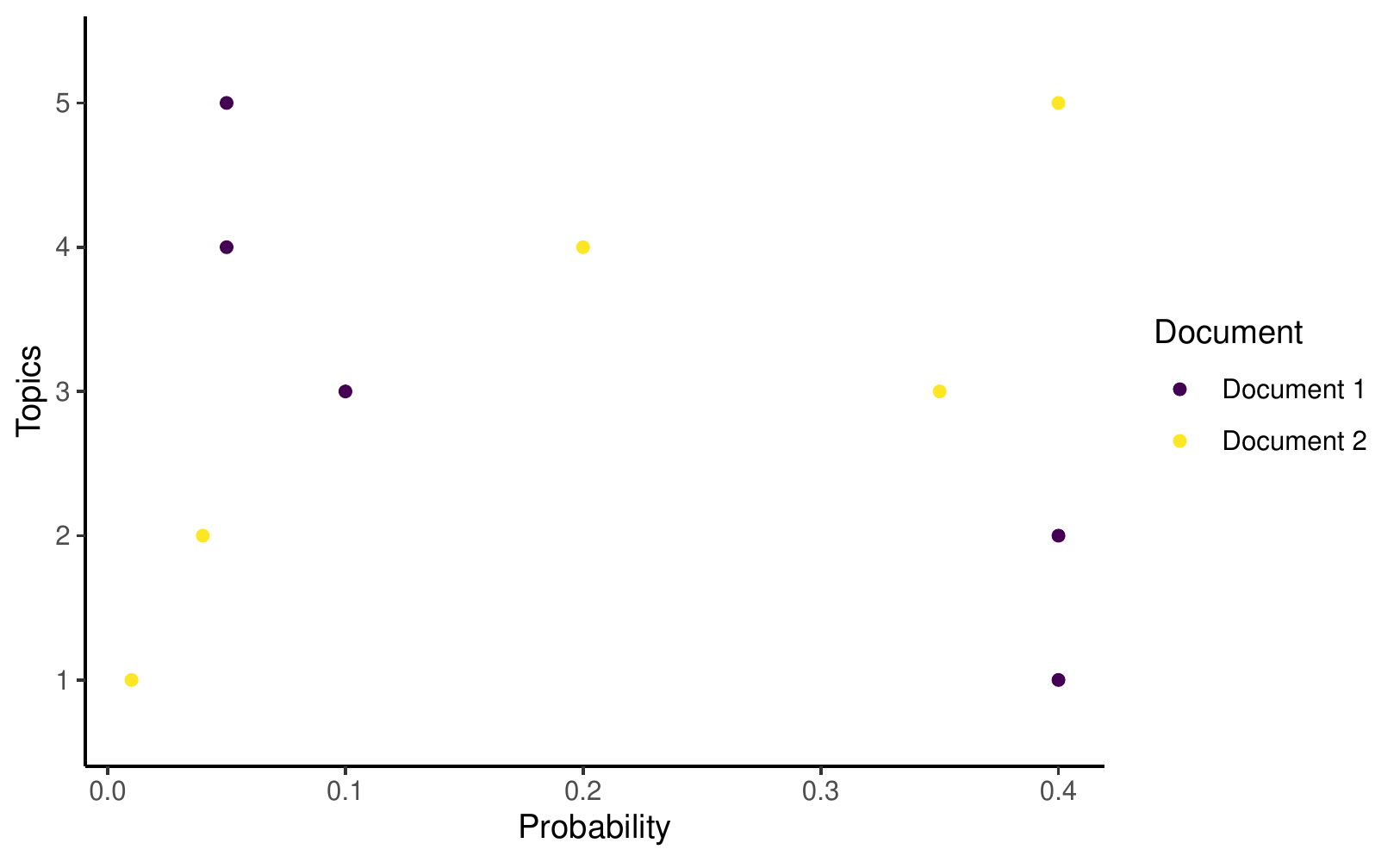}
\caption{\label{fig:topicsoverdocuments}Probability distributions over topics for two documents}
\end{figure}

\begin{figure}
\centering
\includegraphics{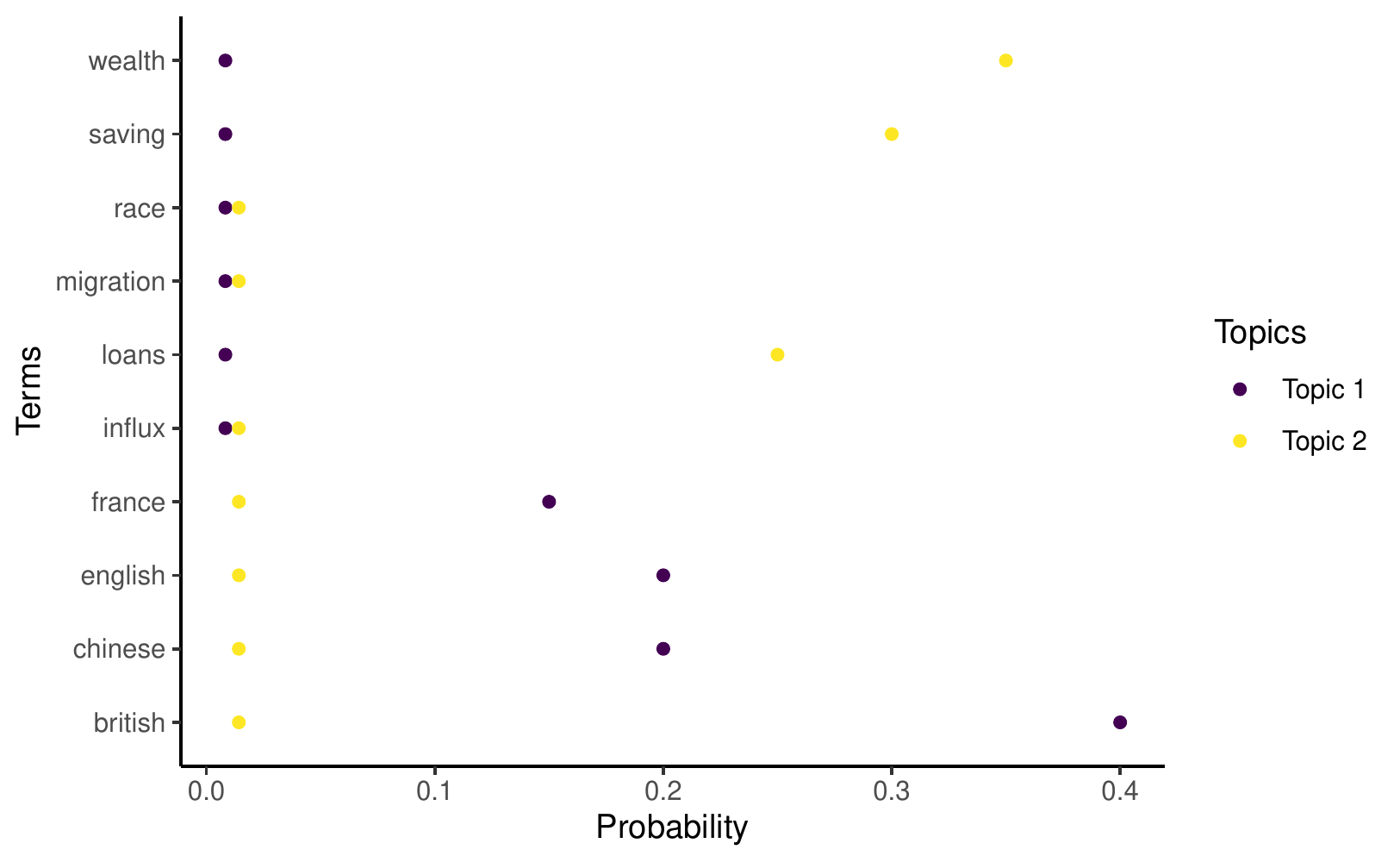}
\caption{\label{fig:topicsoverterms}Probability distributions over terms}
\end{figure}

As an example, Figures \ref{fig:topicsoverdocuments} and \ref{fig:topicsoverterms} illustrate a smaller application with five topics, two documents, and ten terms. In this case, the first document may be comprised mostly of the first few topics; the other document may be mostly about the final few topics (Figure \ref{fig:topicsoverdocuments}).

For instance, if there were ten terms, then one topic could be defined by giving more weight to terms related to immigration; and some other topic may give more weight to terms related to the economy (Figure \ref{fig:topicsoverterms}).

\newpage

\hypertarget{LDAdocgenprocess}{%
\subsection{Document generation process}\label{LDAdocgenprocess}}

Following \citet{BleiLafferty2009}, \citet{blei2012} and \citet{GriffithsSteyvers2004}, the process by which a document is generated is more formally considered to be:

\begin{enumerate}
\def\labelenumi{\arabic{enumi}.}
\tightlist
\item
  There are \(1, 2, \dots, k, \dots, K\) topics and the vocabulary consists of \(1, 2, \dots, V\) terms. For each topic, decide the terms that the topic uses by randomly drawing distributions over the terms. The distribution over the terms for the \(k\)th topic is \(\beta_k\). Typically a topic would be a small number of terms and so the Dirichlet distribution with hyper-parameter \(\boldsymbol{\eta}\) is used: \(\beta_k \sim \mbox{Dirichlet}(\boldsymbol{\eta})\), where \(\boldsymbol{\eta} = (\eta_1, \eta_2, \dots, \eta_{K})\).\footnote{The Dirichlet distribution is a variation of the beta distribution that is commonly used as a prior for categorical and multinomial variables. If there are just two categories, then the Dirichlet and the beta distributions are the same. In the special case of a symmetric Dirichlet distribution, where all elements of \(\eta=1\), it is equivalent to a uniform distribution. If \(\eta<1\), then the distribution is sparse and concentrated on a smaller number of the values, and this number decreases as \(\eta\) decreases. A hyper-parameter is a parameter of a prior distribution.} In practice, a symmetric Dirichlet distribution is usually used, where all elements of \(\boldsymbol{\eta}\) are equal.
\item
  Decide the topics that each document will cover by randomly drawing distributions over the \(K\) topics for each of the \(1, 2, \dots, d, \dots, D\) documents. The topic distributions for the \(d\)th document are \(\theta_d\), and \(\theta_{d,k}\) is the topic distribution for topic \(k\) in document \(d\). Again, the Dirichlet distribution with the hyper-parameter \(0<\alpha<1\) is used here because usually a document would only cover a handful of topics: \(\theta_d \sim \mbox{Dirichlet}(\boldsymbol{\alpha})\). Again, strictly \(\boldsymbol{\alpha}\) is vector of length \(K\) of hyper-parameters and they are usually equal.
\item
  If there are \(1, 2, \dots, n, \dots, N\) terms in the \(d\)th document, then to choose the \(n\)th term, \(w_{d, n}\):

  \begin{enumerate}
  \def\labelenumii{\alph{enumii}.}
  \tightlist
  \item
    Randomly choose a topic for that term \(n\), in that document \(d\), \(z_{d,n}\), from the multinomial distribution over topics in that document, \(z_{d,n} \sim \mbox{Multinomial}(\theta_d)\).
  \item
    Randomly choose a term from the relevant multinomial distribution over the terms for that topic, \(w_{d,n} \sim \mbox{Multinomial}(\beta_{z_{d,n}})\).
  \end{enumerate}
\end{enumerate}

Given this set-up, the joint distribution for the variables is (\citet{blei2012}, p.6):
\[p(\beta_{1:K}, \theta_{1:D}, z_{1:D, 1:N}, w_{1:D, 1:N}) = \prod^{K}_{i=1}p(\beta_i) \prod^{D}_{d=1}p(\theta_d) \left(\prod^N_{n=1}p(z_{d,n}|\theta_d)p\left(w_{d,n}|\beta_{1:K},z_{d,n}\right) \right).\]

Based on this document generation process the analysis problem, discussed next, is to compute a posterior over \(\beta_{1:K}\) and \(\theta_{1:D}\), given \(w_{1:D, 1:N}\). This is intractable directly, but can be approximated (\citet{GriffithsSteyvers2004} and \citet{blei2012}).

After the documents are created, they are all that we have to analyse. The term usage in each document, \(w_{1:D, 1:N}\), is observed, but the topics are hidden, or `latent'. We do not know the topics of each document, nor how terms defined the topics. That is, we do not know the probability distributions of Figures \ref{fig:topicsoverdocuments} or \ref{fig:topicsoverterms}. In a sense we are trying to reverse the document generation process -- we have the terms and we would like to discover the topics.

If the earlier process around how the documents were generated is assumed and we observe the terms in each document, then we can obtain estimates of the topics (\citet{SteyversGriffiths2006}). The outcomes of the LDA process are probability distributions and these define the topics. Each term will be given a probability of being a member of a particular topic, and each document will be given a probability of being about a particular topic. That is, we are trying to calculate the posterior distribution of the topics given the terms observed in each document (\citet{blei2012}, p.~7):
\[p(\beta_{1:K}, \theta_{1:D}, z_{1:D, 1:N} | w_{1:D, 1:N}) = \frac{p\left(\beta_{1:K}, \theta_{1:D}, z_{1:D, 1:N}, w_{1:D, 1:N}\right)}{p(w_{1:D, 1:N})}.\]

Gibbs sampling or the variational expectation-maximization algorithm can be used to approximate the posterior. A summary of these approaches is provided next.

\newpage

\hypertarget{LDAposteriorestimation}{%
\subsection{Posterior estimation}\label{LDAposteriorestimation}}

Following \citet{SteyversGriffiths2006} and \citet{Darling2011}, the Gibbs sampling process attempts to find a topic for a particular term in a particular document, given the topics of all other terms for all other documents. Broadly, it does this by first assigning every term in every document to a random topic, specified by Dirichlet priors with \(\alpha = \frac{50}{K}\) and \(\eta = 0.1\) (\citet{SteyversGriffiths2006} recommends \(\eta = 0.01\)), where \(\alpha\) refers to the distribution over topics and \(\eta\) refers to the distribution over terms (\citet{Grun2011}, p.~7). It then selects a particular term in a particular document and assigns it to a new topic based on the conditional distribution where the topics for all other terms in all documents are taken as given (\citet{Grun2011}, p.~6):
\[p(z_{d, n}=k | w_{1:D, 1:N}, z'_{d, n}) \propto \frac{\lambda'_{n\rightarrow k}+\eta}{\lambda'_{.\rightarrow k}+V\eta} \frac{\lambda'^{(d)}_{n\rightarrow k}+\alpha}{\lambda'^{(d)}_{-i}+K\alpha} \]
where \(z'_{d, n}\) refers to all other topic assignments; \(\lambda'_{n\rightarrow k}\) is a count of how many other times that term has been assigned to topic \(k\); \(\lambda'_{.\rightarrow k}\) is a count of how many other times that any term has been assigned to topic \(k\); \(\lambda'^{(d)}_{n\rightarrow k}\) is a count of how many other times that term has been assigned to topic \(k\) in that particular document; and \(\lambda'^{(d)}_{-i}\) is a count of how many other times that term has been assigned in that document. Once \(z_{d,n}\) has been estimated, then estimates for the distribution of words into topics and topics into documents can be backed out.

This conditional distribution assigns topics depending on how often a term has been assigned to that topic previously, and how common the topic is in that document (\citet{SteyversGriffiths2006}). The initial random allocation of topics means that the results of early passes through the corpus of document are poor, but given enough time the algorithm converges to an appropriate estimate.

The choice of the number of topics, \emph{k}, drives the results and must be specified \emph{a priori}. If there is a strong reason for a particular number, then this can be used. Otherwise, one way to choose an appropriate number is to use cross validation. More detail on this process is provided in the next section.

\newpage

\hypertarget{selecttopicnumber}{%
\subsection{Selection of number of topics}\label{selecttopicnumber}}

The choice of the number of topics to use in a topic model has a substantial effect on the results of the model. For instance, in our topic model, choosing a smaller number of topics, such as 10 or 20 results in a model that is not all that useful because the topics are so broad.

\begin{figure}
\includegraphics[width=1\linewidth]{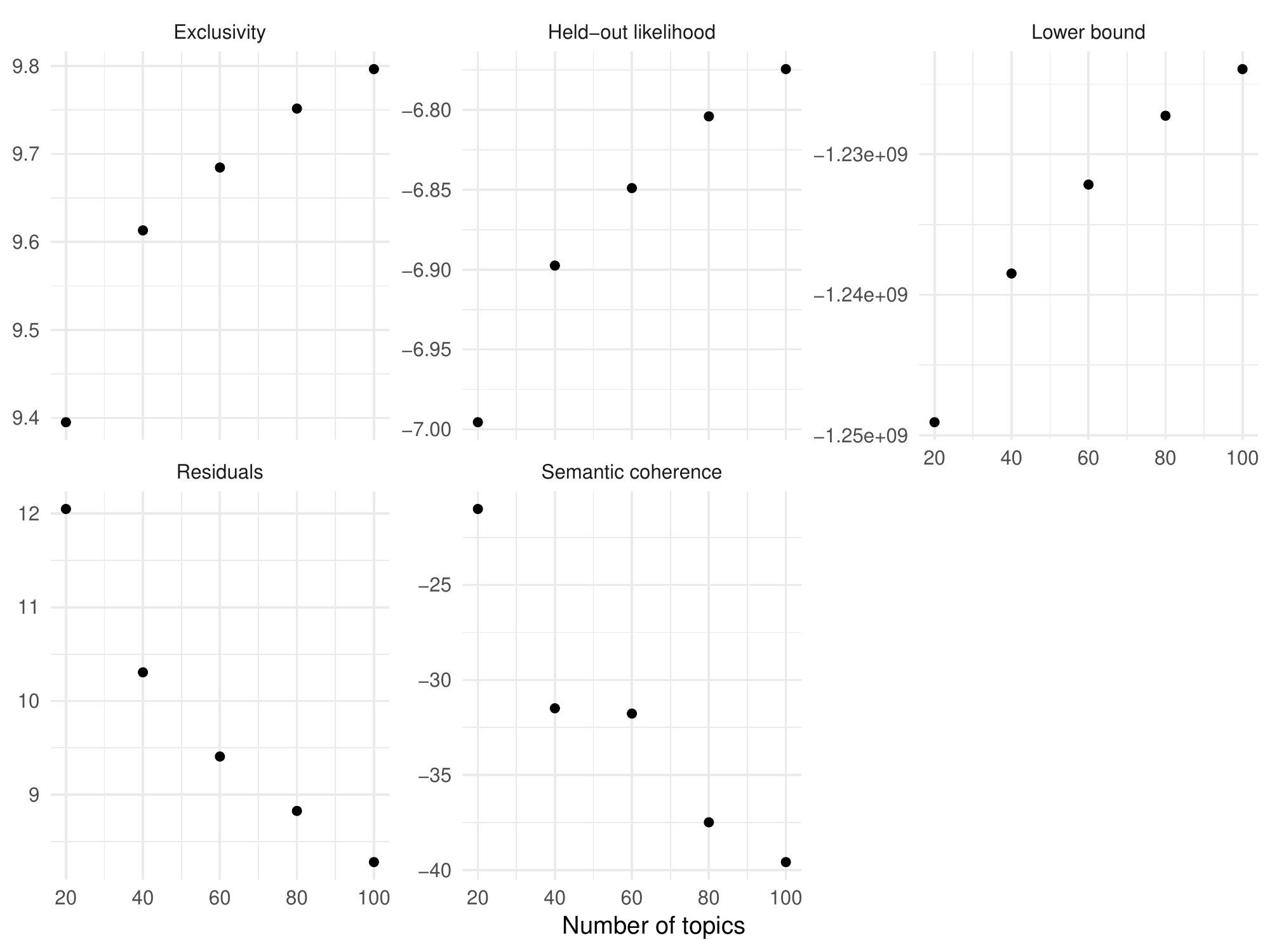} \caption{Model diagnostics}\label{fig:topicsdiagnostics}
\end{figure}

There are a variety of diagnostic measures that can guide the selection of the topics, but there is rarely a clear best choice, especially at a finer level such as choosing between 60 and 65 topics. We found it useful to try a few quite different measures before settling on 80 topics. This provided a balance between being granular enough to be informative---anything less than 40 topics tended to be too broad---yet still being tractable for our analysis model in a reasonable amount of time. In addition to looking at the topics and how they changed over time, diagnostic measures that we considered include the held-out likelihood, the lower bound, residuals, exclusivity. and semantic coherence (Figures \ref{fig:topicsdiagnostics}).\footnote{The code for creating the figures in this section is based on \citet{Silge2018}.}

\begin{figure}
\includegraphics[width=1\linewidth]{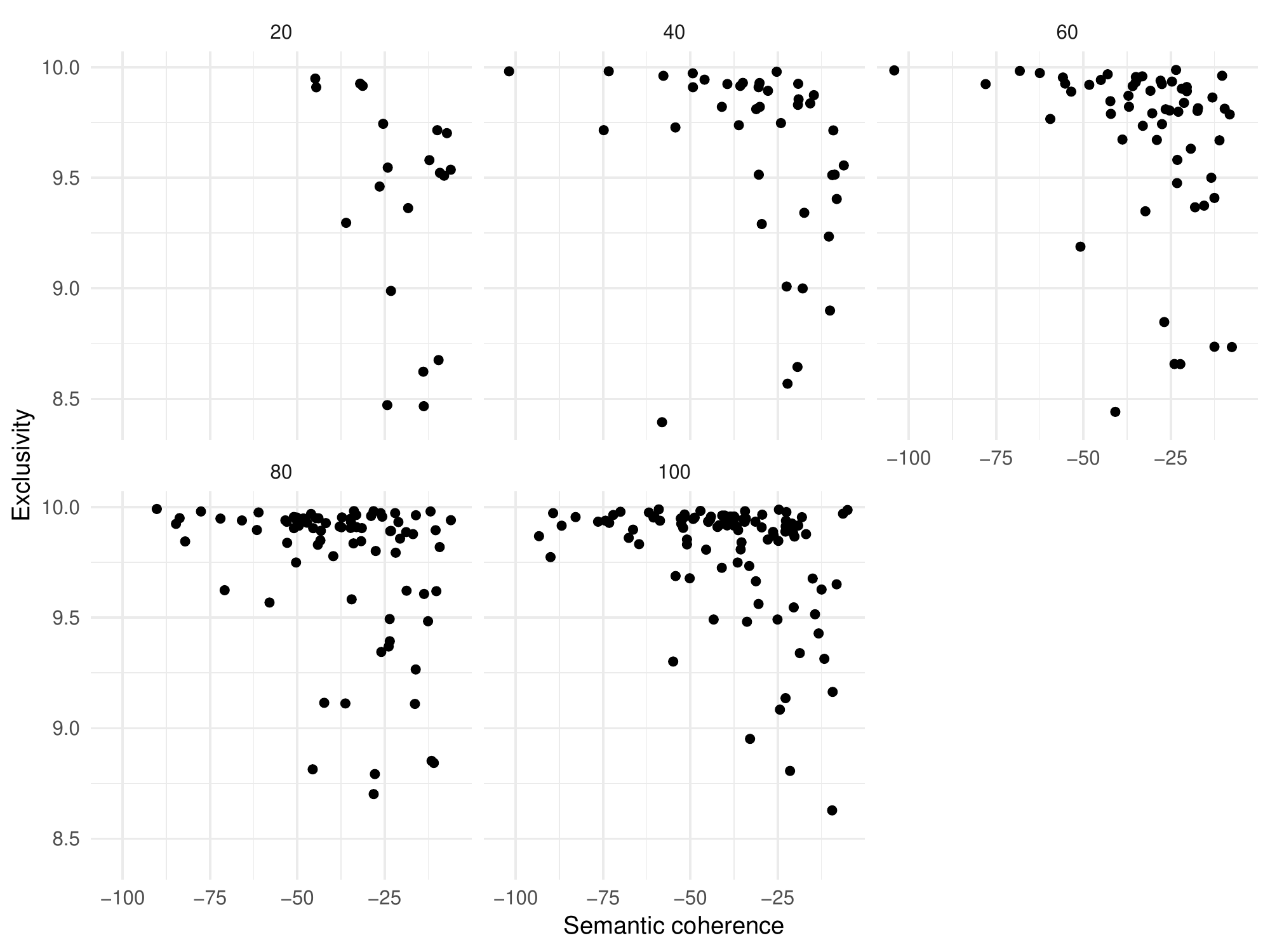} \caption{Exclusitivity compared with semantic coherence}\label{fig:topicsexclusivityvscoherence}
\end{figure}

\citet{RobertsStewartAiroldiRPackage} provides more detail about the diagnostic tests that we use, but we briefly discuss each here. Exclusivity is a measure of how specific words are to particular topics. It looks at the proportion that a word makes up of a particular topic compared with the proportion that word makes up of the other topics. As the number of topics increases we usually expect exclusivity to increase because the topics become more particular. Higher values are better. The held-out likelihood as described by \citet{Wallach2009} takes a test/training approach to estimate the probability of held-out documents given the training documents. Higher values are better. The lower bound gives some indication of whether the model may have multiple modes and hence the end result be sensitive to the starting position (\citet{Roberts2016navigating}). Residuals analysis, \citet{Taddy2012}, compares the theoretical distribution of the variance with the actual distribution. It is a test for over-dispersion of the variance, and if it is found then this can suggest that more topics would be appropriate. Semantic coherence is the trade-off for having topics that are more specific and the subsequent risk that the topics become meaningless. \citet{Mimno2011} define a measure of coherence that is based on ratios of single words compared with pairs of words. The idea is that words that should occur in the one document should be more likely to be in a particular topic than ones that do not occur together. For instance, a topic that has `wine' and `cheese' as highly rated words would score better on their measure than another that contained `cheese' and `mining'. Lower values are better. \citet{RobertsStewartAiroldiRPackage} recommend examining the trade-off between exclusivity and semantic coherence. This suggests that the magnitude of improvement reduces from about 80 topics (Figure \ref{fig:topicsexclusivityvscoherence}).

\hypertarget{correlatedtopicmodelsection}{%
\subsection{Correlated Topic Model}\label{correlatedtopicmodelsection}}

One of the limitations of LDA is that the model assumes that the presence of one topic is not correlated with the presence of another topic. In reality, topics are often related. For instance, in the Hansard context, we may expect topics related to the army to be more commonly found with topics related to the navy, but less commonly with topics related to banking. The goal of the CTM \citep{BleiLafferty2007} is to account for this correlation between topics, in order to produce more realistic and stable topic distributions over time. The models are very similar, and the key difference is the underlying distributions that are drawn from.

As with LDA, the process assumed to generate the documents is the key aspect as this will be reversed to estimate the topics. The document generation process of \citet{Blei2003latent}, discussed in Appendix \ref{LDAdocgenprocess}, is just slightly modified. Specifically, rather than assuming that the distribution of topics in a document, \(\theta_d\), are a draw from a Dirichlet distribution, as in Step 2 of the LDA document generation process detailed in Appendix \ref{LDAdocgenprocess}, CTM assumes:
\[\theta_d \sim \mbox{Logistic Normal}(\mu, \Sigma).\]
That is, the main difference of CTM over LDA is that it replaces the assumption of the Dirichlet distribution with a more flexible logistic multivariate Normal distribution. This distribution can incorporate a covariance structure across the topics. The remainder of the steps of the document generating process are pretty much the same as LDA.

However, the replacement of the Dirichlet distribution with the logistic multivariate Normal distribution adds a level of computational complexity to CTM. The posterior distributions of the parameters of interest (\(\beta_{1:K}, \theta_{1:D}, z_{1:D, 1:N}\)) can no longer be obtained using standard simulation techniques such as Gibbs Sampling. \citet{BleiLafferty2007} develop a fast variational inference procedure for estimating the CTM. CTM itself has also been extended by \citet{RobertsStewartAiroldi2016} as part of their work on Structural Topic Models. The main difference is to add a covariate to \(\mu\) which allows consideration of additional information.

\newpage

\hypertarget{topicmodeleconresultsdiscussion}{%
\section{Topic model outputs - Economics}\label{topicmodeleconresultsdiscussion}}

Although the example topics in the main paper have a military and defence theme, there are other topics that also share a broader theme. An example of these are those to do with economics. For instance Topics 17, 22, 25, 44, 51, 55, 59, 71, 72, 74, 76, and 77.

\begin{figure}
\includegraphics[width=1\linewidth]{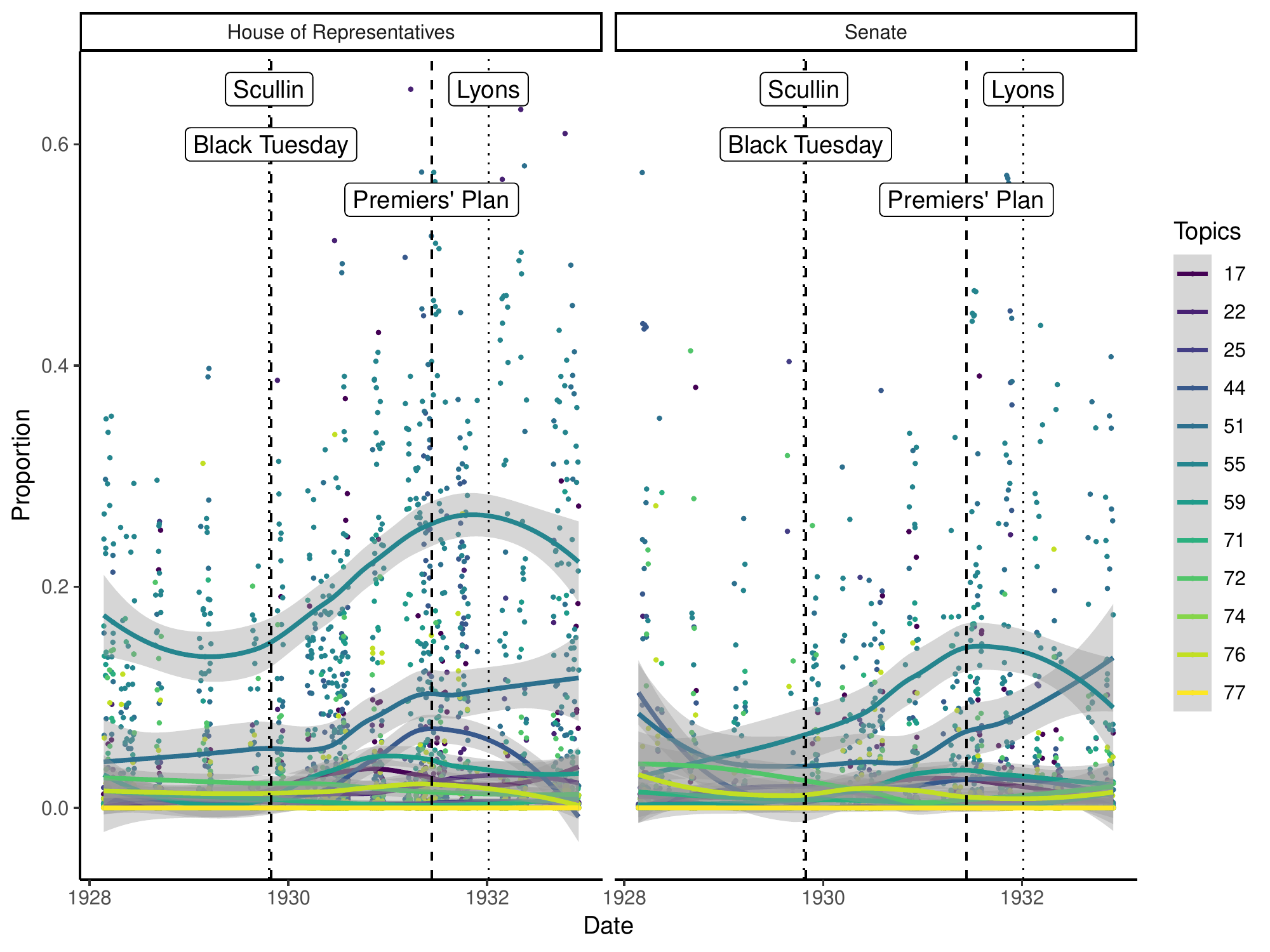} \caption{Economic topic changes around the the Great Depression}\label{fig:greatd}
\end{figure}

In Figure \ref{fig:greatd} we focus on the period around the Great Depression and the Premiers' Plan.

The notable aspect of the response to the 1929 `Black Tuesday' stock market declines was the Premiers' Plan, which appeared newspapers on 11 June 1931. \citet{Copland1934} describes the differences of opinion that occurred in the lead-up to the plan, but Figure \ref{fig:greatd} highlights how new the Scullin Government was when the economic troubles began.

\begin{figure}
\includegraphics[width=1\linewidth]{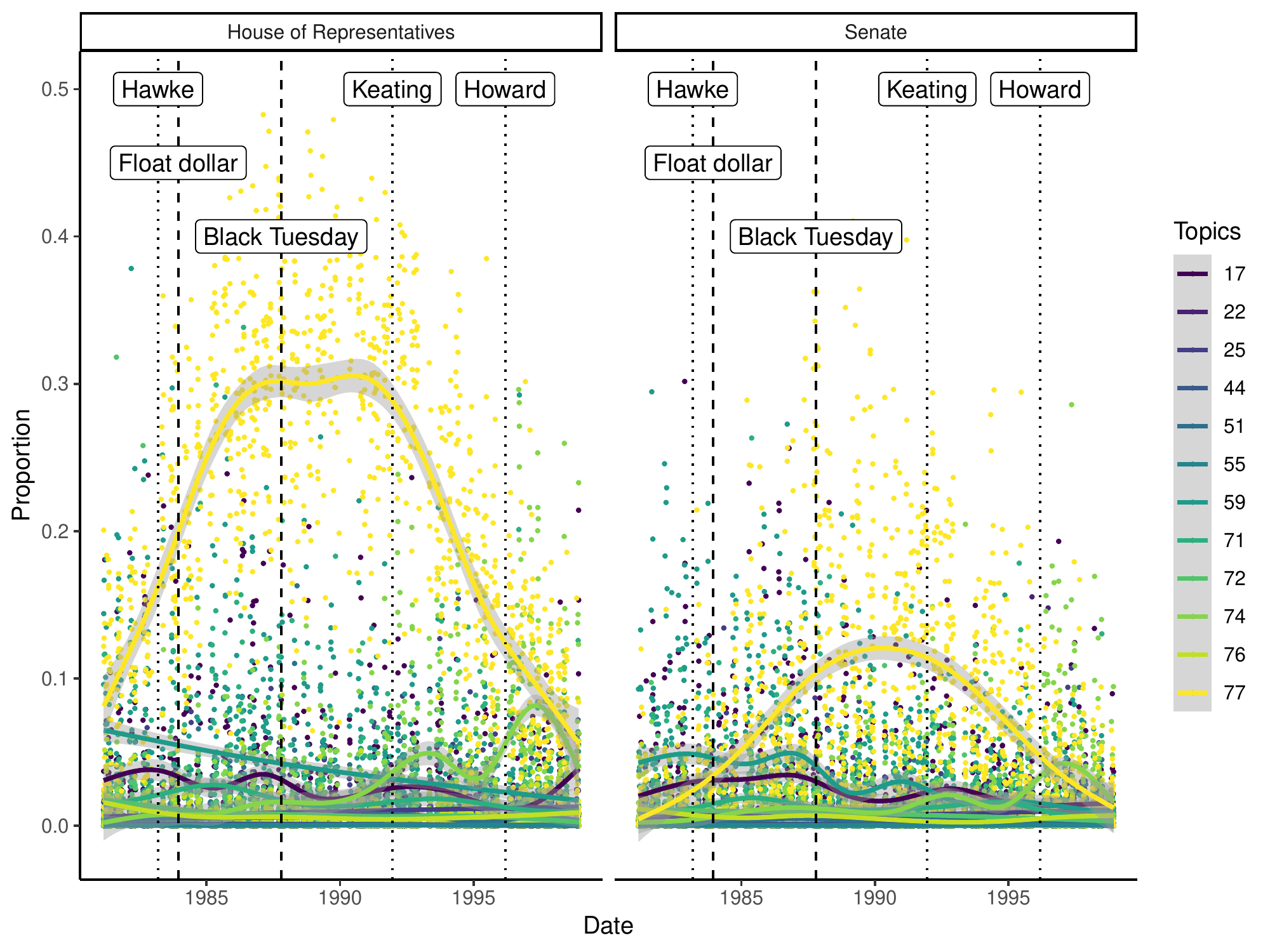} \caption{Economic topic changes during the 1980s and 1990s}\label{fig:theeightsnighties}
\end{figure}

In Figure \ref{fig:theeightsnighties} we focus on the 1980s and 1990s when there was a great deal of economic change.

One notable aspect is the difference between the House of Representatives and the Senate that the proportion of economic topics accounts for. The emphasis that the Hawke/Keating government placed on economic issues comes through clearly given the clear change following Hawke's election.

\begin{figure}
\includegraphics[width=1\linewidth]{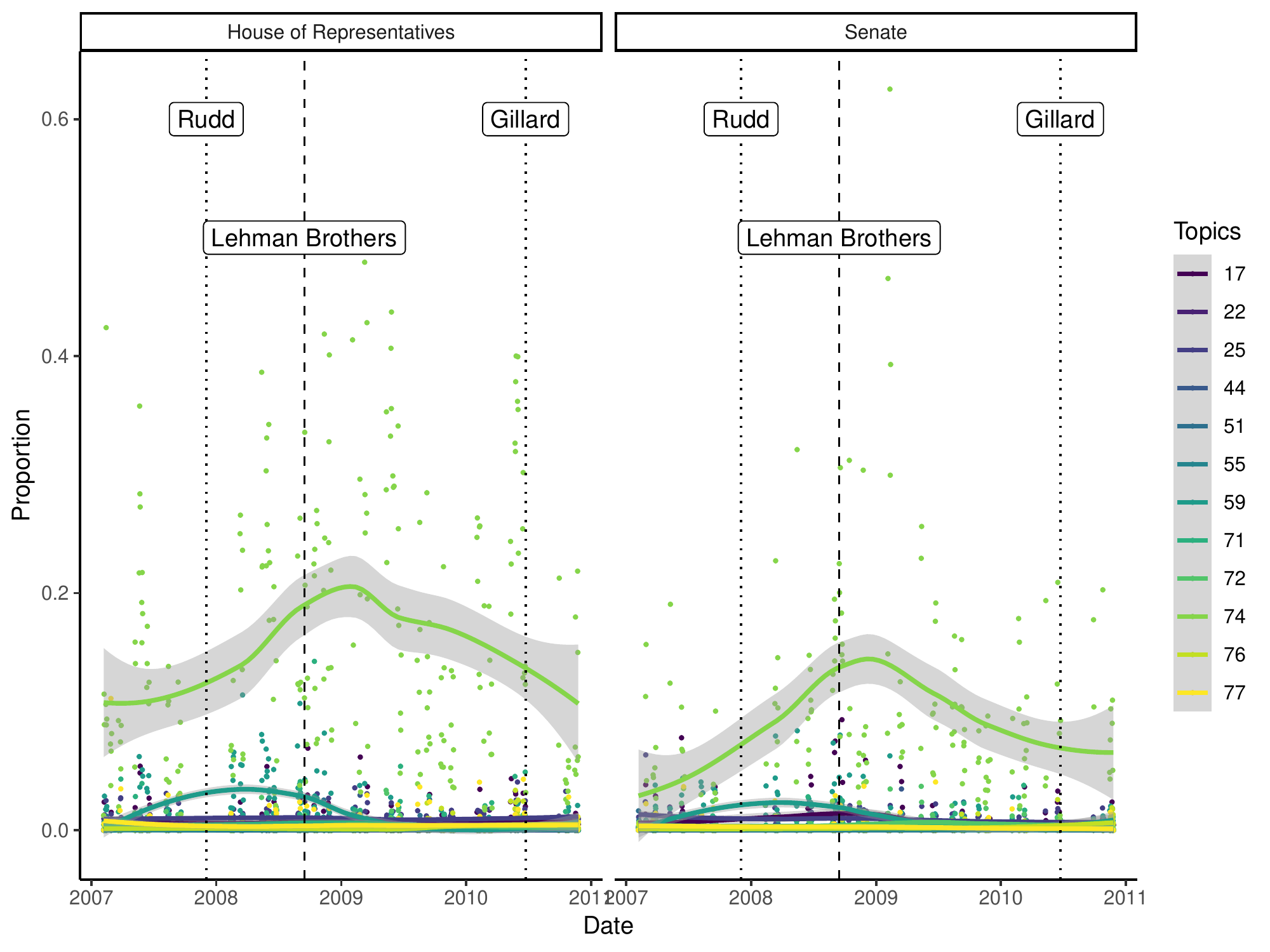} \caption{Economic topic changes around the 2007-08 financial crisis}\label{fig:economicsgfc}
\end{figure}

Finally, in Figure \ref{fig:economicsgfc} we focus on the financial crisis of 2007-08.

It is notable that despite the enormous importance of the financial crisis of 2007-08, topics that are clearly related to economic issues do not dominate the discussion in either house.

\newpage

\hypertarget{eventdetails}{%
\section{Events}\label{eventdetails}}

\hypertarget{fulllistofelections}{%
\subsection{List of Elections}\label{fulllistofelections}}

The first of the two types of events that we consider in this paper is an election (Table \ref{tab:elections}).

\begingroup\fontsize{8}{10}\selectfont

\begin{ThreePartTable}
\begin{TableNotes}[para]
\item \textit{Note: } 
\item This table contains summary information for each prime minister of Australia. HoR and Senate refer to the number of days that each chamber sat for between elections.
\end{TableNotes}
\begin{longtable}[t]{rrlrlrr}
\caption{\label{tab:elections}List of Australian elections}\\
\toprule
Number & Year & Date & Total seats & Winner & HoR & Senate\\
\midrule
1 & 1901 & 1901-03-29 & 75 & Non-labor & 297 & 240\\
2 & 1903 & 1903-12-16 & 75 & Non-labor & 282 & 185\\
3 & 1906 & 1906-12-12 & 75 & Non-labor & 286 & 217\\
4 & 1910 & 1910-04-13 & 75 & Labor & 249 & 167\\
5 & 1913 & 1913-05-31 & 75 & Non-labor & 108 & 59\\
\addlinespace
6 & 1914 & 1914-09-05 & 75 & Labor & 147 & 133\\
7 & 1917 & 1917-05-05 & 75 & Non-labor & 174 & 134\\
8 & 1919 & 1919-12-13 & 75 & Non-labor & 258 & 197\\
9 & 1922 & 1922-12-16 & 75 & Non-labor & 171 & 132\\
10 & 1925 & 1925-11-14 & 75 & Non-labor & 205 & 147\\
\addlinespace
11 & 1928 & 1928-11-17 & 75 & Non-labor & 40 & 29\\
12 & 1929 & 1929-10-12 & 75 & Labor & 206 & 154\\
13 & 1931 & 1931-12-19 & 75 & Non-labor & 155 & 111\\
14 & 1934 & 1934-09-15 & 74 & Non-labor & 169 & 116\\
15 & 1937 & 1937-10-23 & 74 & Non-labor & 153 & 108\\
\addlinespace
16 & 1940 & 1940-09-21 & 74 & Non-labor & 144 & 113\\
17 & 1943 & 1943-08-21 & 74 & Labor & 206 & 132\\
18 & 1946 & 1946-09-28 & 74 & Labor & 278 & 119\\
19 & 1949 & 1949-12-10 & 121 & Non-labor & 107 & 87\\
20 & 1951 & 1951-08-28 & 121 & Non-labor & 173 & 99\\
\addlinespace
21 & 1954 & 1954-05-29 & 121 & Non-labor & 94 & 65\\
22 & 1955 & 1955-12-10 & 122 & Non-labor & 190 & 147\\
23 & 1958 & 1958-11-22 & 122 & Non-labor & 200 & 194\\
24 & 1961 & 1961-12-09 & 122 & Non-labor & 119 & 106\\
25 & 1963 & 1963-11-30 & 122 & Non-labor & 196 & 170\\
\addlinespace
26 & 1966 & 1966-11-26 & 124 & Non-labor & 179 & 185\\
27 & 1969 & 1969-10-25 & 125 & Non-labor & 208 & 209\\
28 & 1972 & 1972-12-02 & 125 & Labor & 97 & 96\\
29 & 1974 & 1974-05-18 & 127 & Labor & 117 & 118\\
30 & 1975 & 1975-12-13 & 127 & Non-labor & 147 & 151\\
\addlinespace
31 & 1977 & 1977-12-10 & 124 & Non-labor & 188 & 196\\
32 & 1980 & 1980-10-18 & 125 & Non-labor & 121 & 152\\
33 & 1983 & 1983-03-05 & 125 & Labor & 101 & 124\\
34 & 1984 & 1984-12-01 & 148 & Labor & 183 & 202\\
35 & 1987 & 1987-07-11 & 148 & Labor & 158 & 215\\
\addlinespace
36 & 1990 & 1990-03-24 & 148 & Labor & 163 & 216\\
37 & 1993 & 1993-03-13 & 147 & Labor & 187 & 213\\
38 & 1996 & 1996-03-02 & 148 & Non-labor & 180 & 196\\
39 & 1998 & 1998-10-03 & 148 & Non-labor & 215 & 214\\
40 & 2001 & 2001-11-10 & 150 & Non-labor & 189 & 161\\
\addlinespace
41 & 2004 & 2004-10-09 & 150 & Non-labor & 196 & 163\\
42 & 2007 & 2007-11-24 & 150 & Labor & 173 & 129\\
43 & 2010 & 2010-08-21 & 150 & Labor & 179 & 155\\
44 & 2013 & 2013-09-07 & 150 & Non-labor & 190 & 153\\
45 & 2016 & 2016-07-02 & 150 & Non-labor & 156 & 137\\
\bottomrule
\insertTableNotes
\end{longtable}
\end{ThreePartTable}
\endgroup{}

Between 1901 and 2018 there are 45 elections, roughly one every two to three years. All of the election periods have a reasonable number of sitting days within them. The fewest was the election on 17 November 1928, with only 40 sitting days in the House of Representatives and 29 in the Senate, as there was another election almost a year later on 12 October 1929.

The number of seats increases from 74 to 121 at the 10 December 1949 election, having been reasonably consistent to that point. Another large increase in the number of seats, this time from 125 to 148, happens at the 1 December 1984 election.

In the first half of our sample especially, the name of the major opposition party changes. For this reason we distinguish between the Labor Party, and the non-Labor Party in terms of who won the election, that is which party was able to form government.

\newpage

\hypertarget{fulllistofgovernments}{%
\subsection{List of Prime Ministers}\label{fulllistofgovernments}}

The second of the two types of events that we consider in this paper is a change in prime minister (Table \ref{tab:governments}).

\begingroup\fontsize{8}{10}\selectfont

\begin{ThreePartTable}
\begin{TableNotes}[para]
\item \textit{Note: } 
\item This table contains summary information for each prime minister of Australia. HoR and Senate refer to the number of days that each chamber sat for while that person was prime minister.
\end{TableNotes}
\begin{longtable}[t]{llllllrr}
\caption{\label{tab:governments}List of Australian prime ministers}\\
\toprule
Government & Prime Minister & Party & Start & End & Died in Office & HoR & Senate\\
\midrule
Barton & Edmund Barton & Protectionist & 1901-01-01 & 1903-09-24 & No & 284 & 228\\
Deakin 1 & Alfred Deakin & Protectionist & 1903-09-24 & 1904-04-27 & No & 31 & 23\\
Watson & Chris Watson & Labour & 1904-04-27 & 1904-08-18 & No & 47 & 18\\
Reid & George Reid & Free Trade & 1904-08-18 & 1905-07-05 & No & 60 & 37\\
Deakin 2 & Alfred Deakin & Protectionist & 1905-07-05 & 1908-11-13 & No & 334 & 255\\
\addlinespace
Fisher 1 & Andrew Fisher & Labour & 1908-11-13 & 1909-06-02 & No & 15 & 14\\
Deakin 3 & Alfred Deakin & Commonwealth Liberal & 1909-06-02 & 1910-04-29 & No & 94 & 67\\
Fisher 2 & Andrew Fisher & Labor & 1910-04-29 & 1913-06-24 & No & 249 & 167\\
Cook & Joseph Cook & Commonwealth Liberal & 1913-06-24 & 1914-09-17 & No & 108 & 59\\
Fisher 3 & Andrew Fisher & Labor & 1914-09-17 & 1915-10-27 & No & 90 & 77\\
\addlinespace
Hughes & Billy Hughes & Labor National Labor and Nationalist & 1915-10-27 & 1923-02-09 & No & 489 & 387\\
Bruce & Stanley Bruce & Nationalist (Coalition) & 1923-02-09 & 1929-10-22 & No & 416 & 308\\
Scullin & James Scullin & Labor & 1929-10-22 & 1932-01-06 & No & 206 & 154\\
Lyons & Joseph Lyons & United Australia (Coalition) & 1932-01-06 & 1939-04-07 & Yes & 396 & 279\\
Page & Earle Page & Country (Coalition) & 1939-04-07 & 1939-04-26 & No & 2 & 0\\
\addlinespace
Menzies 1 & Robert Menzies & United Australia (Coalition) & 1939-04-26 & 1941-08-28 & No & 118 & 87\\
Fadden & Arthur Fadden & Country (Coalition) & 1941-08-28 & 1941-10-07 & No & 8 & 6\\
Curtin & John Curtin & Labor & 1941-10-07 & 1945-07-05 & Yes & 223 & 153\\
Forde & Frank Forde & Labor & 1945-07-06 & 1945-07-13 & No & 1 & 1\\
Chifley & Ben Chifley & Labor & 1945-07-13 & 1949-12-19 & No & 357 & 173\\
\addlinespace
Menzies 2 & Robert Menzies & Liberal (Coalition) & 1949-12-19 & 1966-01-26 & No & 1024 & 822\\
Holt & Harold Holt & Liberal (Coalition) & 1966-01-26 & 1967-12-19 & Yes & 117 & 111\\
McEwen & John McEwen & Country (Coalition) & 1967-12-19 & 1968-01-10 & No & 0 & 0\\
Gorton & John Gorton & Liberal (Coalition) & 1968-01-10 & 1971-03-10 & No & 200 & 201\\
McMahon & William McMahon & Liberal (Coalition) & 1971-03-10 & 1972-12-05 & No & 125 & 128\\
\addlinespace
Whitlam & Gough Whitlam & Labor & 1972-12-05 & 1975-11-11 & No & 213 & 213\\
Fraser & Malcolm Fraser & Liberal (Coalition) & 1975-11-11 & 1983-03-11 & No & 457 & 500\\
Hawke & Bob Hawke & Labor & 1983-03-11 & 1991-12-20 & No & 546 & 681\\
Keating & Paul Keating & Labor & 1991-12-20 & 1996-03-11 & No & 246 & 289\\
Howard & John Howard & Liberal (Coalition) & 1996-03-11 & 2007-12-03 & No & 780 & 734\\
\addlinespace
Rudd 1 & Kevin Rudd & Labor & 2007-12-03 & 2010-06-24 & No & 172 & 128\\
Gillard & Julia Gillard & Labor & 2010-06-24 & 2013-06-27 & No & 179 & 154\\
Rudd 2 & Kevin Rudd & Labor & 2013-06-27 & 2013-09-18 & No & 1 & 2\\
Abbott & Tony Abbott & Liberal (Coalition) & 2013-09-18 & 2015-09-15 & No & 143 & 115\\
Turnbull & Malcolm Turnbull & Liberal (Coalition) & 2015-09-15 & 2018-08-24 & No & 179 & 151\\
\addlinespace
Morrison & Scott Morrison & Liberal (Coalition) & 2018-08-24 & - & - & 24 & 24\\
\bottomrule
\insertTableNotes
\end{longtable}
\end{ThreePartTable}
\endgroup{}

Between 1901 and 2018 30 different people have been prime minister. Four of those---Deakin, Fisher, Menzies and Rudd---returned as prime minister at least once after being replaced, meaning Scott Morrison defines the 36th prime ministerial period.

Three people have died while prime minister: Lyons in 1939, Curtin in 1945, and Holt in 1967. Their respective immediate successors were only prime minister for a short period. As such, we do not consider them when determining the neighbouring prime minister. Specifically, we compare: the first Menzies term with Lyons instead of Page; Chifley with Curtin instead of with Forde; and Gorton with Holt instead of with McEwan.

The other prime ministerial period that we do not consider in this paper is the second Rudd term. This is because it only contained a few sitting days. That is, we compare Tony Abbot with Julia Gillard.

\newpage

\newpage
\singlespacing 
\renewcommand\refname{References}
\bibliography{bibliography.bib}

\end{document}